\documentstyle[fleqn,aps,epsfig]{revtex}
\begin{document}

\title{Series expansion studies of random sequential adsorption 
with diffusional relaxation}

\author{Chee Kwan Gan and Jian-Sheng Wang}
\address{Department of Computational Science, \\
National University of Singapore, Singapore 119260,\\
Republic of Singapore.
}
\date{29 July, 1996}
\maketitle

\begin{abstract}
We obtain long series (28 terms or more) for the coverage 
(occupation fraction) $\theta$, in powers of time $t$ for 
two models of random sequential adsorption with diffusional
relaxation using an efficient algorithm developed by
the authors.  Three different kinds of analyses of the series are
performed for a wide range of $\gamma$, the rate of diffusion of
the adsorbed particles, to investigate the power law approach of 
$\theta$ at large times.
We find that the
primitive series expansions in time $t$ for $\theta$ capture
rich short and intermediate time kinetics of the systems very
well. However, we see that the series are still not long enough to
extract the kinetics at large times for general $\gamma$.
We have performed extensive
computer simulations employing an efficient event-driven algorithm 
to confirm the $t^{-1/2}$ saturation approach of $\theta$ at
large times for both models, as well as to investigate the
short and intermediate time behaviors of the systems.

PACS number(s): 05.70.Ln, 82.20.Mj, 75.40.Mg, 05.50.+q
\end{abstract}

\section*{Introduction}
Random sequential adsorption (RSA) \cite{Evans-93} is an irreversible process
which particles are deposited randomly and consecutively on a surface.
The depositing particles, represented by hard-core extended objects,
satisfy the excluded volume condition where they are not allowed to overlap. 
The exclusion of certain regions
for further deposition attempts due to the adsorbed particles leads
to a dominant infinite-memory correlation effect where the system
approaches partially covered, fully blocked stage at large times. 
However, this picture is altered when the diffusional relaxation
is introduced \cite{Privman-Nielaba-92,Nielaba-Privman-92,%
Wang-group-93}.
Privman and Nielaba \cite{Privman-Nielaba-92} have shown 
that the effect of added diffusional relaxation in the deposition of 
dimer on a 1D lattice substrate is to allow the full, saturation coverage
via a $\sim t^{-1/2}$ power law at large times, preceded
by a mean-field crossover regime with the 
intermediate $\sim t^{-1}$ behavior for fast diffusion.

Series expansion is one of the powerful analytical methods in the
RSA studies \cite{Baram-Kutasov-89,Dickman-Wang-Jensen-91,%
Oliveira-group-92,Bonnier-group-93,%
Baram-Fixman-95,Gan-Wang-96}. Long series in powers
of time $t$ have been obtained,
reminiscent of series expansions in equilibrium statistical mechanics,
by using a computer \cite{Martin-74}.
Recently, the authors \cite{Gan-Wang-96}
have proposed an efficient algorithm for generating
long series for the coverage $\theta$
in powers of time $t$ based on the hierarchical rate equations.

The present work is to study the time-dependent quantity
$\theta$ for one-dimensional models of RSA with 
diffusional relaxation, both analytically and numerically.
It will be seen that even though relatively
long series have been obtained, we are still unable to extract
the kinetics of the systems at large times for general $\gamma$
due to long, rich transient crossover regime that the series must
describe.
Extensive computer simulations are performed to confirm the $t^{-1/2}$ 
power law approach of $\theta$, where we have employed an 
efficient event-driven algorithm.
The remainder of this paper is organized as follows.
Section~\ref{sec:model} introduces two related models. 
Details of series expansion are explained
in Section~\ref{sec:series_expansion}.
Analyses of the series can be found in
Section~\ref{sec:series_analysis}. Monte Carlo results
are presented in Section~\ref{sec:monte_carlo} and finally
Section~\ref{sec:conclusion}
contains the summary and conclusions.

\section{The models}
\label{sec:model}
Two models have been studied in this work. We start with
an initially empty, infinite linear lattice. Dimers are dropped randomly 
and sequentially 
at a rate of $k$ per lattice site
per unit time, onto the lattice. Hereafter we set $k$ equal to
unity without loss of generality. If the chosen two
neighbor sites are unoccupied, the dimer is adsorbed on the lattice. If
one of the chosen sites is occupied, the adsorption attempt is rejected.
One of the simplest possibilities of diffusional relaxation in this dimer
adsorption process is that
the adsorbed dimer
is permitted to hop either to left or right by one lattice
constant at a diffusion rate $\gamma$ from the original dimer position,
provided that the diffusion attempt does not violate the excluded volume
condition.
This model
has been initiated and studied by 
Privman and Nielaba \cite{Privman-Nielaba-92}.
We refer this model as the dimer RSA with dimer diffusion or
diffusive dimer model.

A second possibility is that an adsorbed dimer 
is allowed to dissociate into two independent
monomers; each monomer can diffuse to one of its nearest neighbor sites
with a diffusion rate $\gamma$, provided that
the diffusion attempt does not violate the excluded volume condition.
This model bears a strong resemblance to the 
former model and is exactly solvable when $\gamma = 1/2$
\cite{Grynberg-Stinchcombe-95}. 
We refer this model as the dimer RSA with monomer diffusion or 
diffusive monomer model.
Interestingly enough,
the special case of the diffusive monomer problem 
with $\gamma = 1/2$ can be mapped to 
the diffusion-limited process
\begin{equation}
{\cal A} + {\cal A} \rightarrow \hbox{inert},
\end{equation} which is known as one-species annihilation
process \cite{Avraham-group-90}.
This model has been solved exactly by a number of researchers
\cite{Lushnikov-87,Spouge-88,Balding-group-88}.
We observe that when $\gamma = 1/2$, the effect of a dimer 
deposition attempt in the diffusive monomer model corresponds to
two diffusion attempts of $\cal A$ in an adjacent pair of $\cal A$
of the ${\cal A} + {\cal A} \rightarrow \hbox{inert}$ process.
The time-dependent quantity coverage 
$\theta(t)$ (fraction of occupied sites)
for the diffusive monomer model with $\gamma = 1/2$ 
is given by
\begin{eqnarray}
\theta(t) = 1- \exp(-2t)I_0(2t),
\end{eqnarray}
where $I_n(z) $ is the modified Bessel function of integer order $n$.
\unitlength=6pt
\def\ock{\circle{0.6}}
\def\xck{\circle*{0.6}}
\def\o{\begin{picture}(1,1)(-0.5,-0.5) 
\put(0,0){\ock}
\end{picture}}
\def\x{\begin{picture}(1,1)(-0.5,-0.5) 
\put(0,0){\xck}
\end{picture}}
\def\oo{\begin{picture}(2,1)(-0.5,-0.5)  
\put(0,0){\ock}
\put(1,0){\ock}
\end{picture}}
\def\ooo{\begin{picture}(3,1)(-0.5,-0.5) 
\put(0,0){\ock}
\put(1,0){\ock}
\put(2,0){\ock}
\end{picture}}
\def\oox{\begin{picture}(3,1)(-0.5,-0.5)
\put(0,0){\ock}
\put(1,0){\ock}
\put(2,0){\xck}
\end{picture}}
\def\oxo{\begin{picture}(3,1)(-0.5,-0.5)
\put(0,0){\ock}
\put(1,0){\xck}
\put(2,0){\ock}
\end{picture}}
\def\oooo{\begin{picture}(4,1)(-0.5,-0.5)
\put(0,0){\ock}
\put(1,0){\ock}
\put(2,0){\ock}
\put(3,0){\ock}
\end{picture}}
\def\ooox{\begin{picture}(4,1)(-0.5,-0.5)
\put(0,0){\ock}
\put(1,0){\ock}
\put(2,0){\ock}
\put(3,0){\xck}
\end{picture}}
\def\ooxo{\begin{picture}(4,1)(-0.5,-0.5)
\put(0,0){\ock}
\put(1,0){\ock}
\put(2,0){\xck}
\put(3,0){\ock}
\end{picture}}
\def\ooxx{\begin{picture}(4,1)(-0.5,-0.5) 
\put(0,0){\ock}
\put(1,0){\ock}
\put(2,0){\xck}
\put(3,0){\xck}
\end{picture}}
\def\oxxo{\begin{picture}(4,1)(-0.5,-0.5) 
\put(0,0){\ock}
\put(1,0){\xck}
\put(2,0){\xck}
\put(3,0){\ock}
\end{picture}}
\def\oooxx{\begin{picture}(5,1)(-0.5,-0.5) 
\put(0,0){\ock}
\put(1,0){\ock}
\put(2,0){\ock}
\put(3,0){\xck}
\put(4,0){\xck}
\end{picture}}
\def\ooxxo{\begin{picture}(5,1)(-0.5,-0.5) 
\put(0,0){\ock}
\put(1,0){\ock}
\put(2,0){\xck}
\put(3,0){\xck}
\put(4,0){\ock}
\end{picture}}

\section{Series expansions}
\label{sec:series_expansion}
To illustrate how series expansions are performed, we note
that the first few rate equations for the dimer
and monomer diffusive models are 
\begin{eqnarray}
{dP(\o) \over dt} & = & -2 P(\oo), \label{eq:1stdimer}\\
{dP(\oo) \over dt} & = & -P(\oo) - 2P(\ooo) - 
		   2 \gamma P(\ooxx) + 2 \gamma P(\oxxo),
\label{eq:2nddimer} \\
{dP(\ooo)\over dt} & = & -2P(\ooo) - 2P(\oooo) - 2\gamma P(\oooxx) 
+2 \gamma P(\ooxxo), \\
\cdots \nonumber
\end{eqnarray}
and 
\begin{eqnarray}
{dP(\o) \over dt} & = & -2 P(\oo), \\
{dP(\oo) \over dt} & = & -P(\oo) - 2P(\ooo) - 2\gamma P(\oox)
  + 2\gamma P(\oxo), \\
{dP(\ooo) \over dt} & = & -2P(\ooo) - 2P(\oooo) 
- 2\gamma P(\ooox) + 2\gamma P(\ooxo), \\
{dP(\oox)\over dt} & = & -P(\oox) + P(\oooo) - \gamma P(\oox)
 + \gamma P(\oxo) - \nonumber \\
		  &  & \gamma P(\ooxo) + \gamma P(\ooox), \\
\cdots \nonumber
\end{eqnarray}
respectively, where $P(C)$ denotes the probability
of finding a configuration $C$ of sites specified empty
`$\o$' or filled `$\x$'. Unspecified sites can be occupied or
empty. Here we have taken into account the 
symmetries of a configuration under all lattice group operations.
For the one-dimensional configurations, we just need to 
consider the reflection operation only. 

Let $C_o$ denote a particular configuration of interest, and
$P_{{C_o}} \equiv P(C_o)$ the associated configuration probability.
$P_{C_o}$ is expected to be a well behaved function of time $t$,
so one can
obtain the Taylor series expansion with the expansion
point at $t = 0$, 
$P_{C_o}(t) = \sum\limits_{n=0}^{\infty} {P_{C_o}}^{(n)}t^n/n! $, with
the $n$th {\it derivative} of $P_{C_o}$ given by
\begin{equation}
{P_{C_0}}^{(n)} = \left. {d^n P_{C_0}(t) \over d t^n}\right|_{t=0}.
\end{equation}

Let $G_i$ denote the set of new configurations generated in
the calculation of the $i$th derivative of $P_{C_0}$, and 
$G_i^j$ the corresponding $j$th
derivatives of the set of configurations.
We observe that $G_0^{n-1}$, $G_1^{n-2}$, 
$\ldots$, $G_{n-1}^0$ (determined at the $(n-1)$th derivative), 
$G_0^{n-2}$, $G_1^{n-3}$, $\ldots$,
$G_{n-2}^0$ (determined at the $(n-2)$th derivative), $\ldots$,
$G_0^0$ are predetermined before calculating the $n$th derivative
of $P_{C_o}$. In the calculation of $n$th derivative of $P_{C_o}$,
we determine systematically $G_0^n$, $G_1^{n-1}$, $\ldots$, $G_{n-1}^1$,
$G_{n}^0$, by recursive use of rate equations. This algorithm is 
efficient since each value
in $G_i^{n-i}$, $ 0 \le i \le n$ and 
the rate equation for a configuration $C$ is generated once only.
However, this algorithm consumes the memory quickly as a result of 
storage of intermediate results.

The computation of the expansion coefficients makes use of the 
isomorphism between a lattice configuration and
its binary representation if we map an occupied
(empty) site to 1 (0). 
The data structures used to represent Eq.~(\ref{eq:1stdimer})
and Eq.~(\ref{eq:2nddimer}) are depicted in Fig.~\ref{fig:flow}.
A node for a configuration $C$ is characterized by its four
components; (i) the representation of $C$ 
in the computer, (ii) a pointer
to the derivatives of $P_C$, ${P_C}^{(n)}$ for $n = 1, 2, 3, \ldots$,
(iii) the highest order of derivative $h$ of $P_C$ 
obtained so far, and (iv) a
pointer to a linked list of nodes of configurations
(`children') which appear in the
right hand side of the rate equation for $P_C$.
The linked list contains the associated coefficients for each
`child'.
The variable $h$ is used so that we know the values of
${P_C}^{(n)}$ where $ 1 \le n \le h$ have already been calculated 
and can be retrieved when needed.
All pointers to the configuration nodes generated during the
enumeration process are stored in a hash table or a binary tree
to allow efficient checking of the existence of 
any configuration. Use of the algorithm and data structures
allows us to obtain 
coefficients up to $t^{31}$ and $t^{27}$
(presented in Appendix
\ref{appendix:series_coefficient})
for $P(\o,t)$ of the diffusive dimer and monomer models, respectively.

\section{Analyses of series}
\label{sec:series_analysis}
Analytically, we are interested in confirming
the power law approach
of $t^{-1/2}$ of the coverage $\theta$ at large times for both
diffusive dimer and monomer models through the 
unbiased and biased analyses of the series. 
The unbiased analysis does not fix the saturation coverage of the system,
while the biased analysis assumes the saturation coverage
to be the value 1.
For the unbiased analysis, let $\theta(t) = 1 - P(\o, t)$ be 
the time dependent coverage. Let assume that at very large times $t$
the coverage $\theta$ satisfies the equation
\begin{eqnarray}
\theta(t) = \theta_c - \frac{A(t)}{t^{\delta}},
\label{eq:power_law}
\end{eqnarray}
where $\theta_c$ is the saturation coverage.
$A(t)$ is assumed to be a
function of $t$, which tends to a constant value as $t\rightarrow\infty$,
and $\delta$ is the exponent that characterizes the saturation
approach (we expect to obtain $\delta = 1/2$ from the
analysis of the series for all $\gamma$ values).
Writing $t = A(t)^{1/\delta}(\theta_c - \theta)^{-1/\delta}$,
we see that if we perform a DLog Pad\'e 
\cite{Baker-61,Hunter-Baker-73,Baker-Hunter-73} analysis
to the inverted series $t= t(\theta)$, where
\begin{eqnarray}
\frac{d}{d\theta} \log t(\theta) =
\frac{1}{\delta}\frac{d}{d\theta} \log A(t) - 
\frac{1}{\delta}\cdot\frac{1}{\theta-\theta_c},
\end{eqnarray}
then the power law of Eq.~(\ref{eq:power_law})
implies a simple, isolated pole of $\theta_c$ with
an associated residue of $-1/\delta$. Fig.~\ref{fig:explode} shows
the plot of the inverted series $t$ versus $\theta$ 
for the 28-term series with $\gamma = 1/2$ for the monomer diffusive
model. 

For the diffusive dimer problem, 
the closest real pole to the value 1 (the expected saturation coverage)
for [16,15], [15,16], [15,15], [16,14], and [14,16]
Pad\'e approximants
are shown in
Fig.~\ref{fig:dimer_theta}, with the corresponding saturation
exponents $\delta$ displayed in Fig.~\ref{fig:dimer_delta}.
Similarly we form
[14,13], [13,14], [13,13], [14,12], [12,14] Pad\'e approximants
for the diffusive monomer problem, where the results are 
displayed in Fig.~\ref{fig:monomer_theta}
and Fig.~\ref{fig:monomer_delta}.
Comparing the graphs for these two models, the diffusive dimer
series give a better convergence of $\theta_c$ and $\delta$ against 
$\gamma$ than that for the diffusive monomer
series generally, presumably due to the fact that the
coefficients of the series of $P(\o, t)$
alternate in signs in the former model.
For small $\gamma$ 
values ($\gamma < 5$), the estimates for $\theta_c$ and
$\delta$ are 
unstable --- different 
Pad\'e approximants do not agree with
one another. 
The series with $\gamma = 0$ describes a pure lattice RSA
behavior \cite{Dickman-Wang-Jensen-91}, where 
the system approaches the jamming coverage exponentially.
Hence we expect the confirmation for 
power law of Eq.~(\ref{eq:power_law}) is interfered by the 
exponential behavior of the series when $\gamma$ is small.
For $\gamma > 5$, there are physically
favorable estimates for $\theta_c$ and $\delta $ where 
$\theta_c = 1.00 \pm 0.05$ and 
$\delta = 1.0\pm 0.1$ for $10 < \gamma < 20$, for both models.
These results are the manifestations of 
the transient regime of $t^{-1}$ approach to saturation.

The distribution plot of the poles and zeros in the vicinity of
$(1, 0) $ is displayed in Fig.~\ref{fig:zeropole} for the 
$[14, 13]$ Pad\'e approximant for the 28-term series with $\gamma
= 1/2$ for the diffusive monomer model. We see that the
real pole closest to $(1,0)$ is not distinguished and isolated
from the nearby poles and zeros. 
This explains 
the difficulty of unbiased analysis that
the intermediate crossover effect masks the power law approach
at late stages.

We also perform biased analyses for the series. This
series analysis have been used by Jensen and Dickman 
\cite{Jensen-Dickman-93}  to extract
critical exponents from series in powers of time $t$. We define
the $F$-transform of $f(t)$ by
\begin{equation}
F[f(t)] = t \frac{d}{dt} \ln f.
\end{equation}
If $f \sim A t^{-\alpha}$ for some constant
$A$, then $F(t) \rightarrow \alpha$ as 
$t \rightarrow \infty$. We consider the exponential 
transformation
\begin{equation}
\label{eq:transform}
z = \frac{1-e^{-bt}}{b},
\end{equation}
which proved to be very useful in the analysis of RSA series 
\cite{Dickman-Wang-Jensen-91,Gan-Wang-96,Jensen-Dickman-93}.
This transformation involves a parameter $b$ which cannot be
fixed a priori is then followed by the construction of various orders
of Pad\'e approximants to the $z$-series.
Crossing region is then searched for in the
graphs of $\alpha$ versus $b$, the transformation parameter.

To illustrate this biased analysis, 
we take the saturation coverage $\theta_c$ to be 1 and choose
$f(t)$ to be $P(\o,t)= \theta_c - \theta(t)$. Since we expect
$P(\o,t) \sim t^{-1/2}$ for large times $t$,
specifically we have formed [14,13],
[13,14], [13,13], [14,12], [12,14] Pad\'e approximants to the
$z$-series for the 28-term series with $\gamma = 1/2$ for the
diffusive monomer model.  We find that 
the estimates for $\delta$ is $0.5061(5)$, 
for $ 0.45 < b < 0.50 $, 
as we can see from
Fig.~\ref{fig:biased_exact}. 
Thus the exact analytical function of $\exp(-2t)I_0(2t)$
serves as a useful guide of
this analysis, where the exponent deviates from 
the value $1/2$ by only about 1\%.

Given a value of $\gamma$, we obtain the corresponding
estimates of $\delta$
from the first convergence of all Pad\'e
approximants by locating the crossing region. 
The results of $\delta$ estimates for several values of $\gamma$ 
are presented in Table~\ref{tab:allexp}.
The corresponding uncertainties for $\delta$
which reflect the variation of $\delta$ over a range of $b$ are 
shown in the same
table.
For the diffusive dimer model,
we have formed [16,15], [15,16], [15,15], [16,14], [14,16],
[15,14], and [14,15] Pad\'e approximants to the $z$-series.
The corresponding graphs are displayed in 
Fig.~\ref{fig:allexp}. It is seen that for small values of
$\gamma$, 
we obtain small estimates of $\delta$, while 
for large $\gamma$, $ \delta \rightarrow 1$, 
suggesting the approach to
the limiting saturation is via a mean-field like result, i.e. 
the $t^{-1}$ power law. 
Hence we see that even though the exponential transformation
Eq.~(\ref{eq:transform}) works well for the exact series of 
of diffusive monomer model when $\gamma = 1/2$, its use for general
$\gamma$ is not very appropriate. We have also tried the 
transformation
$ z = 1 - (1 + bt)^{-1/2}$ to the series for both diffusive
models but the convergence is rather poor.

We have tried and used a third 
method of extracting the saturation exponent $\delta$.
If we assume that for large enough times $t$, 
the saturation coverage $\theta$ assumes a power law
\begin{equation}
1 - \theta \propto t^{-\delta},
\end{equation}
then we expect a plot of $d\ln(1-\theta)/d\ln(t)$ versus 
$t$ or $\log_{10}(t)$ should give a plateau 
of constant $-\delta$ values. By forming [14,13], [13,14],
[13,13] Pad\'e approximants to the $d\ln(1-\theta)/d\ln(t)$ of the 28-term
series for the diffusive monomer model with $\gamma = 1/2$,
we observe from Fig.~\ref{fig:illus} that the agreement between
different Pad\'e estimates and the exact solution is excellent 
for $\log_{10}(t)$ up to around $0.9$.
For diffusive dimer problem, three Pad\'e approximants of
[16,15], [15,16], and [15,15] are formed. 
The plots of $d\ln(1-\theta)/d\ln(t)$ versus
$\log_{10}(t)$ for the diffusive dimer and monomer models,
shown in Fig.~\ref{fig:dimerwing} and Fig.~\ref{fig:monomwing}, 
respectively,
are obtained by taking the average of the 3 different Pad\'e
estimates.
The graphs end before the difference 
between at least a pair of Pad\'e estimates is more
that 0.001. 
The last estimates 
in Fig.~\ref{fig:dimerwing} and Fig.~\ref{fig:monomwing}
are taken as the estimates for $\delta$ and
they are listed in the last two columns of Table~\ref{tab:allexp}.
These estimates for $\delta$ are plotted 
in the same graph for the $F$-transformed 
analysis for comparisons (Fig.~\ref{fig:allexp}). 
It is seen that our last method of extracting the
saturation exponents appears to be better than
the $F$-transform analysis since
it yields almost about the same estimates for $\delta$.
It does not involve any
transformation which is not known in advance that will yield
consistent results \cite{Jensen-Dickman-93}. Looking 
at the ends of the curves in Fig.~\ref{fig:dimerwing} and
Fig.~\ref{fig:monomwing}, we are certain that the power law
regime is
still not reached since the $\delta$ estimates do not seem
to converge to
a constant value,
except the case when $\gamma = 1/2$ for the 
diffusive monomer model. 
From this we know that
our estimates for $\delta$ 
do not describe the true power law approach at large times $t$.
Such information cannot be found in the $F$-transform analysis.
We note that our last method of analyzing the series 
is easy to use compared to the $F$-transform analysis.

\section{Monte Carlo simulations}
\label{sec:monte_carlo}
To study the short and large time
behaviors of the coverage, we have performed extensive and
exhaustive simulations for the diffusive dimer and monomer models. 
For both models, we take an initially empty linear lattice with 
$N = 20000$ sites with periodic boundary conditions so that
the finite-size effects can be ignored. In each Monte Carlo step,
a pair of adjacent sites is chosen randomly.
The type of attempted process is then decided: deposition
with probability 
$p$, where $ 0 < p \le 1$, or diffusion with
probability $(1-p)$.
In the case of
the deposition attempts, if any one of the chosen sites is occupied,
the deposition attempt is rejected (unsuccessful attempt), else
the adsorption attempt is accepted.
In the case of diffusion, yet another decision
is made either to move right or left, with equal probability.  If
the selected decision is diffusion to the right,
we check the selected pair of sites are occupied
and
its right nearest neighbor site is unoccupied,
then the dimer is moved by one lattice constant to the right. The 
left-diffusion attempts are treated similarly.
In contrast to the diffusive dimer model, the diffusive monomer
model allows monomers to move by one lattice constant.

We define one time unit interval ($\Delta t = 1$) 
to be during which a deposition
attempt is performed for each lattice site. Thus for
$N$-site lattice, one unit time corresponds to $N$ deposition
attempts, on average. The diffusion rate $\gamma$ relative to the 
deposition rate, is then $\gamma = (1-p)/2p$. 

Straightforward simulation procedure, as described above,
encounters a serious drawback in which at late stages, 
most adsorption and diffusion attempts are rejected.
In order to study the behavior of the system at large times, we
have used an event-driven algorithm to speed up the dynamics 
of the simulations \cite{Brosilow-group-91,Wang-94}. 
Let $q$ be the probability that we can make a successful
move, then the probability that the first $(i-1)$th trials is unsuccessful,
and the $i$th trial is successful is 
\begin{eqnarray}
P_i = q (1-q)^{i-1}, \hbox{\ \ \ } i = 1, 2, 3, \ldots
\label{eq:event_driven}
\end{eqnarray}
If we restrict all trials to be coming from the successful ones,
then two consecutive trials are in fact separated by a random variable $i$ 
in Eq.~(\ref{eq:event_driven}). This distribution can be generated
by
\begin{eqnarray}
i = \left\lfloor\frac{\ln \xi}{\ln(1-q)}\right\rfloor + 1,
\end{eqnarray} 
where $\xi$ is a uniformly distributed random number between 0 and 1.
In employing this method, we have to keep and update an 
active list of successful
moves/attempts, where from its length we can evaluate $q$ at any instance. 

Simulations are performed on a cluster of fast workstations. Our numerical
results are obtained for $\gamma  = $ 0.05, 0.10, 0.20, $\ldots$, 
and 6.40, for  $t$ up to $2^{20}$.
Each data set is averaged over 500 runs, and the longest run
take about 150 CPU hours on a HP712/60. The coverage (fraction of 
occupied sites), $\theta(t)$, is plotted in Fig.~\ref{fig:dimer_phase}
and Fig.~\ref{fig:monom_phase} for
the diffusive dimer and monomer models, respectively.
We have also performed the simulation at $\gamma = 1/2$ for
the diffusive monomer model in order to compare the simulation
results with the exact
results. It is seen that the agreement between them are so good
that actually an overlapping line is observed in
Fig.~\ref{fig:monom_phase}. 
For $\gamma = 0$, we have the exact solution \cite{Dickman-Wang-Jensen-91}
\begin{eqnarray}
\theta(t) = \frac{1-\exp(-2(1-\exp(-t) ))}{2}.
\end{eqnarray}
For extremely fast diffusion case, i.e., $\gamma = \infty$, 
exact results have been 
obtained, where
\begin{eqnarray}
t(\theta) = \frac{1}{4}\bigl(\frac{1}{1-\theta} - 1\bigr) 
- \frac{1}{4} \ln(1-\theta),
\label{eq:dimer_fast_diff}
\end{eqnarray}
for the diffusive dimer model \cite{Privman-Barma-92} and 
\begin{eqnarray}
t(\theta) = \frac{1}{2}\bigl(\frac{1}{1-\theta} - 1\bigr).
\label{eq:monom_fast_diff}
\end{eqnarray}
for diffusive monomer model.
The approach of $(\theta_c - \theta) \sim t^{-1} $ at all times is obvious 
for these extremely fast diffusion models. We have included the 
lines of slope
$-1/2$ to indicate the $t^{-1/2}$ power law clearly. 
It is seen from Fig.~\ref{fig:dimer_phase} and
Fig.~\ref{fig:monom_phase} that 
for $\gamma  \ge 3.20$, the system takes a very long time
($t \approx 10^{4}$) before it can 
enter the final $t^{-1/2}$ regime. This explain why we have
difficulty in extracting the actual power law
approach from the primitive expansion in time $t$.

To further confirm that the saturation approach indeed follows a power
law, we have used a scaling analysis.
For large times $t$, let us assume that $(1-\theta)$ has the following
scaling form 
\begin{equation}
\label{eq:functional}
1-\theta = (\gamma t)^{-1/2}G(\gamma^b/t) ,
\end{equation}
where $G$ is a scaling function and $b$ is a constant to be determined.
Eq. (\ref{eq:functional}) requires that $G(u)$ tends to a
constant when $u$ tends to 0 \cite{Privman-private}.
This is required because for large times $t$,
$(1-\theta) \sim t^{-1/2}$. Let further assume that for large $u$, $G(u)
\sim u^z$ for some constant $z$.
For extremely large $\gamma$, we have $ (1 -\theta) \sim t^{-1}$
(see Eqs. (\ref{eq:dimer_fast_diff}) and
(\ref{eq:monom_fast_diff}))
hence $z = 1/2$ and $b = 1$.
Writing Eq. (\ref{eq:functional}) as
\begin{eqnarray*}
1-\theta & = &(1/\gamma)(\gamma/t)^{1/2}G(\gamma/t) \\
	 & \equiv & (1/\gamma)F(t/\gamma),
\end{eqnarray*}
we then make log-log plots of $\gamma (1-\theta)$ versus
$t/\gamma$ for the diffusive dimer and monomer models as shown in
Figs. \ref{fig:dimerscale} and \ref{fig:monomscale},
respectively. It is seen that the data collapse into single
curves at large times $t$. Our numerical data confirm not only
the power law decay but also a scaling behavior for $(1-\theta)$.

\section{Conclusions}
\label{sec:conclusion}
By using an efficient algorithm based on the hierarchical rate
equations, relatively long series are
obtained for two models of 1-d RSA with diffusional relaxation. 
Analyses of series are performed, but it is seen that 
even though the series are long, we only manage to extract 
the behaviors of the systems up to intermediate times only.
To study the power law of a system at large times $t$,
we see that a series which exhibits a continuous crossover
behavior in its short and intermediate times ought to be long
enough so that various orders of Pad\'e approximants can still
converge in the power law regime.
Using these long series, we find that the analysis 
of series based on the ratio method by Song and Poland
\cite{Song-Poland-92} was not useful. Specifically,
for the diffusive monomer series when $\gamma = 1/2$ (which
corresponds to the A + A $\rightarrow$ 0 process in 
Song and Poland's work), we obtain the saturation
exponent (where they have used the symbol
$\nu$) $\delta$ $=$ $2.0$, $1.2$, 
$0.895$, $0.729$, $0.624$, $0.551$, $0.498$, $0.458$,
$0.426$, $0.401$, $0.380$, $0.363$, $0.351$, $\ldots$, 
which is not seen
to be converging towards the expected value of $1/2$.

We have also performed extensive computer simulations using
an efficient event-driven algorithm, where it
allows us to use and simulate a larger system to much
larger times $t$ than it was done previously on a supercomputer
\cite{Privman-Nielaba-92}. The $t^{-1/2}$ power law approach
of $\theta$ to its saturation is confirmed numerically 
at large times $t$. 

\section*{Acknowledgement}
This work was supported in part by
an Academic Research Grant RP950601 of National University of
Singapore. Part of the calculations were performed on the
facilities of the Computation Center of the Institute of Physical
and Chemical Research, Japan. We would like to thank
V. Privman for pointing out Ref. \cite{Grynberg-Stinchcombe-95}
and useful discussion on how to analyse the series. 
We also appreciate one of the referees for
suggesting us to make a scaling behavior study.

\pagebreak
\bibliographystyle{plain}

\pagebreak
\begin{figure}
\caption{The data structures used to represent the rate equations
Eq.~(\ref{eq:1stdimer}) and Eq.~(\ref{eq:2nddimer}). The 
first field of the node associated with `o' is the representation
of the pattern `o'. The
second field
points to its first four derivatives (i.e. -2,
6, -22, 94). The third field is the highest derivative, $h$
obtained so far for $P(o)$; in this case $h$ is 4. 
The rate equation is represented in 
the fourth field.
The rate equation for the configuration `oo' involves four
configurations, one of them is `oo' itself.}
\label{fig:flow}
\end{figure}

\begin{figure}
\caption{The plot of the inverted series of time $t$ 
versus coverage $\theta$ for the
28-term series with $\gamma =1/2$ for the diffusive monomer
model.}
\label{fig:explode}
\end{figure}

\begin{figure}
\caption{The results for the saturation coverage $\theta_c$
as a function of the rate of diffusion $\gamma$
obtained from the DLog Pad\'e analysis of the inverted series of 
$t = t(\theta)$ with [16,15], [15,16], [15,15],
[16,14], [14,16] Pad\'e approximants for 
the diffusive dimer model. 
The saturation coverage estimates are very close to 1.}
\label{fig:dimer_theta}
\end{figure}

\begin{figure}
\caption{The results for the saturation exponent,
$\delta$ as a function of $\gamma$ obtained
from DLog Pad\'e analysis of the inverted series 
$t = t(\theta)$ for the diffusive dimer model. These estimates
for $\delta$ are deduced from the residues 
associated with the poles in Fig.~\ref{fig:dimer_theta}.}
\label{fig:dimer_delta}
\end{figure}

\begin{figure}
\caption{The results for the saturation coverage $\theta_c$
as a function of the rate of diffusion $\gamma$
obtained from the DLog Pad\'e analysis of the inverted series of
$t = t(\theta)$ with [14,13], [13,14], [13,13],
[14,12], [12,14] Pad\'e approximants for
the diffusive monomer model.}
\label{fig:monomer_theta}
\end{figure}

\begin{figure}
\caption{The results for the saturation exponent,
$\delta$ as a function of $\gamma$ obtained 
from DLog Pad\'e analysis of the inverted series
$t = t(\theta)$ for the diffusive monomer model. These estimates
for $\delta$ are deduced from the residues 
associated with the poles in
Fig.~\ref{fig:monomer_theta}.}
\label{fig:monomer_delta}
\end{figure}

\begin{figure}
\caption{The distribution of zeros and poles in the vicinity of
(1, 0) for the $[14,13]$
Pad\'e approximant for the 28-term series with $\gamma =1/2$ for
the diffusive monomer model. A circle (cross) denotes a pole
(zero).}
\label{fig:zeropole}
\end{figure}

\begin{figure}
\caption{Pad\'e approximant estimates for the exponent $\delta$,
as a function the transformation parameter $b$,
derived from the 
$F$-transform analysis of the 28-term series with 
$\gamma = 1/2$ 
for the monomer diffusive model.}
\label{fig:biased_exact}
\end{figure}

\begin{figure}
\caption{
The plot of the values in Table~\ref{tab:allexp}. The diamonds
and 
the squares denote the estimates of $\delta$ from the $F$-transform
analysis for the diffusive dimer and monomer models, respectively. 
The crosses and the circles correspond to the 
estimates of $\delta$ from the $d\ln(1-\theta)/d\ln(t)$ versus
$\log_{10}(t)$ type analysis, for the 
diffusive dimer and monomer models, respectively.
The error bars are smaller
than the symbols and hence they are not displayed.}
\label{fig:allexp}
\end{figure}

\begin{figure}
\caption{The analysis based on the $d\ln(1-\theta)/d\ln(t)$
versus $\log_{10}(t)$ plot
for [14,13], [13,14], and [13,13] Pad\'e approximants. 
We see that the agreement between three Pad\'e estimates and the exact
results is excellent for $\log_{10}(t)$ up to around 0.9.}
\label{fig:illus}
\end{figure}

\begin{figure}
\caption{The plot of $d\ln(1-\theta)/d\ln(t)$ versus
$\log_{10}(t)$ for the diffusive dimer problem. All curves end
before the difference between
at least a pair of estimates from 
[16,15], [15,16], [15,15] Pad\'e approximants is
greater than 0.001. The ends of curves for $\gamma$ = 0.1,
0.2, $\ldots$, 1.0, 1.5, 2.0, \ldots, 5.0, are displayed in the
downward direction.}
\label{fig:dimerwing}
\end{figure}

\begin{figure}
\caption{The plot of $d\ln(1-\theta)/d\ln(t)$ versus
$\log_{10}(t)$ for the diffusive monomer problem. All curves end
before the difference between
at least a pair of estimates from 
[14,13], [13,14], [13,13] Pad\'e approximants
is greater than 0.001.The ends of curves for 
$\gamma = $ 0.1, 0.2, $\ldots$, 1.0, 1.5, 2.0, \ldots, 5.0, are
displayed in the downward direction.}
\label{fig:monomwing}
\end{figure}

\begin{figure}
\caption{Monte Carlo simulation results for the 
diffusive dimer model. 
The sequence of $\gamma$ for curves between the exact curves for
$\gamma = 0$ and $\gamma = \infty$, in the downward direction, is
0.05, 0.10, 0.20, \ldots, 6.40. The line of slope $-1/2$
shows the $t^{-1/2}$ approach at large times $t$.}
\label{fig:dimer_phase}
\end{figure}

\begin{figure}
\caption{Monte Carlo simulation results for the 
diffusive monomer model. 
The sequence of $\gamma$ for curves between
the exact curves for $\gamma = 0 $ and $\gamma = \infty$,
in the downward direction, is 0.05, 0.10, 0.20, 0.40, 0.50, 0.80, 1.60, 3.20
and 6.40. Notice that the simulation results for $\gamma
= 1/2$ and the exact results agree with each other extremely well
that only one line is seen. The line of slope $-1/2$
shows the $t^{-1/2}$ approach at large times $t$.} 
\label{fig:monom_phase}
\end{figure}

\begin{figure}
\caption{The scaling plot for the diffusive dimer problem. The
sequence of $\gamma$, in the downward direction, is 6.40, 3.20,
1.60, $\ldots$, 0.05. The line of slope $-1/2$ is included to
indicate the power law clearly.}
\label{fig:dimerscale}
\end{figure}

\begin{figure}
\caption{The scaling plot for the diffusive monomer problem. The
sequence of $\gamma$, in the downward direction, is 6.40, 3.20,
1.60, $\ldots$, 0.05. The line of slope $-1/2$ is included to
indicate the power law clearly.}
\label{fig:monomscale}
\end{figure}

\begin{table}
\caption{The $F$-transform analysis gives
the second and fourth columns which show the 
estimates of $\delta$ deduced from the crossing regions of
the graphs of $\delta$ versus the transformation parameter $b$,
taken in the range indicated in the third and fifth columns.
The last two columns show the results obtained from the
$d\ln(1-\theta)/d\ln(t)$ versus $\log_{10}(t)$ type analysis.}
\label{tab:allexp}
\end{table}

\clearpage
\begin{figure}
\begin{minipage}[t]{16cm}
\begin{center}
TABLE 1.
\vskip1cm
\begin{tabular}{ccccccc}
& \multicolumn{2}{c}{dimer} & \multicolumn{2}{c}{monomer} \\
$\gamma$ & $\delta$  & $b$ & $\delta$ & $b$ & dimer & monomer \\
\hline
0.1	& 0.431(3)	& 0.28 -- 0.33 & 0.269(4) & 0.65 -- 0.70 
& 0.407(1) & 	0.256(1) \\
0.2	& 0.499(2)	& 0.30 -- 0.35 & 0.395(4) & 0.30 -- 0.35 
& 0.510(1) & 	0.375(1) \\
0.3	& 0.542(2)	& 0.35 -- 0.40 & 0.441(2) & 0.35 -- 0.40 
& 0.566(1) & 	0.437(1) \\
0.4	& 0.573(3)	& 0.35 -- 0.40 & 0.47788(4) & 0.45 -- 0.50
& 0.603(1) & 	0.479(1) \\
0.5	& 0.599(4)	& 0.36 -- 0.41 & 0.5061(5) & 0.45 -- 0.50 
& 0.618(1) & 	0.508(1) \\
0.6 	& 0.623(5)	& 0.38 -- 0.43 & 0.535(1) & 0.75 -- 0.80
& 0.662(1) &	0.539(1) \\
0.7	& 0.649(4)	& 0.44 -- 0.49 & 0.557(2) & 0.85 -- 0.90 
& 0.682(1) &	0.562(1) \\
0.8	& 0.669(5)	& 0.47 -- 0.52 & 0.576(1) & 0.90 -- 0.95
& 0.716(1) &	0.580(1) \\
0.9	& 0.685(5)	& 0.48 -- 0.53 & 0.5915(6) & 0.95 -- 1.00
& 0.673(1) &	0.595(1) \\
1.0	& 0.702(5)	& 0.51 -- 0.56 & 0.6050(6) & 1.00 -- 1.05
& 0.684(1) &	0.608(1) \\
1.5	& 0.759(5)	& 0.57 -- 0.62 & 0.6534(3) & 1.25 -- 1.30
& 0.810(1) &	0.650(1) \\
2.0	& 0.792(5)	& 0.58 -- 0.63 & 0.6822(4) & 0.95 -- 1.00
& 0.830(1) &	0.671(1) \\
2.5 	& 0.830(3)	& 0.73 -- 0.78 & 0.7041(4) & 0.75 -- 0.80
& 0.836(1) &	0.693(1) \\
3.0	& 0.854(4)	& 0.80 -- 0.85 & 0.718(2) & 0.80 -- 0.85
& 0.859(1) & 	0.708(1) \\
3.5 	& 0.87(1)	& 0.81 -- 0.86 & 0.730(3) & 0.80 -- 0.85
& 0.865(1) & 	0.704(1) \\
4.0	& 0.885(5)	& 0.88 -- 0.93 & 0.742(4) & 0.75 -- 0.80
& 0.875(1) &	0.705(1) \\
4.5	& 0.898(4)	& 0.90 -- 0.95 & 0.746(4) & 0.85 -- 0.90
& 0.884(1) &	0.710(1) \\
5.0	& 0.907(3)	& 0.95 -- 1.00 & 0.752(4) & 0.85 -- 0.90
& 0.906(1) & 	0.707(1) \\
\end{tabular}
\end{center}
\end{minipage}
\end{figure}

\clearpage
\begin{figure}
\begin{minipage}[t]{16cm}
\begin{center}
FIG.1.
\vskip1cm
\epsfig{file=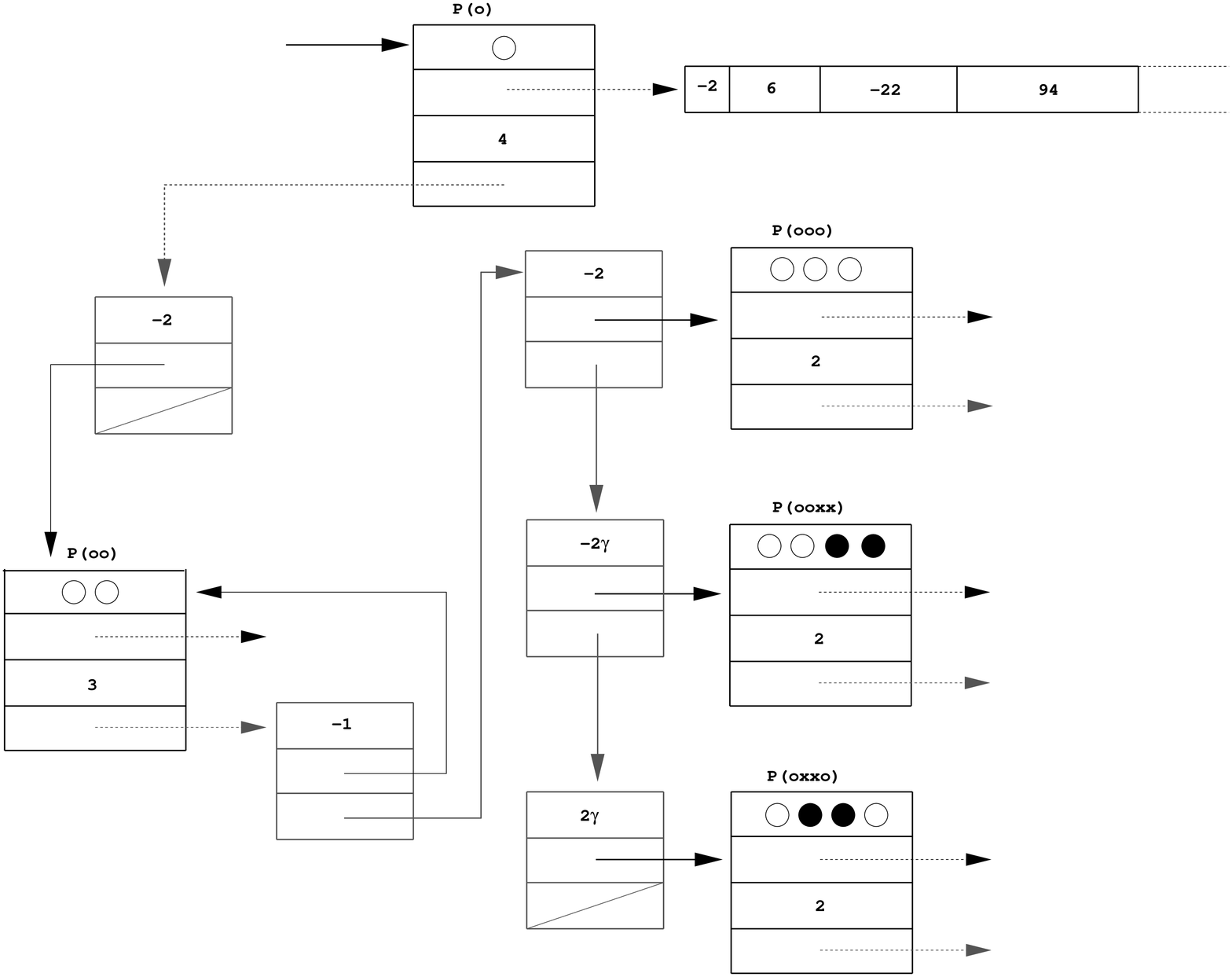,height=14cm,angle=90}
\end{center}
\end{minipage}
\end{figure}

\clearpage
\begin{figure}
\begin{minipage}[t]{16cm}
\begin{center}
FIG.2.
\vskip1cm
\epsfig{file=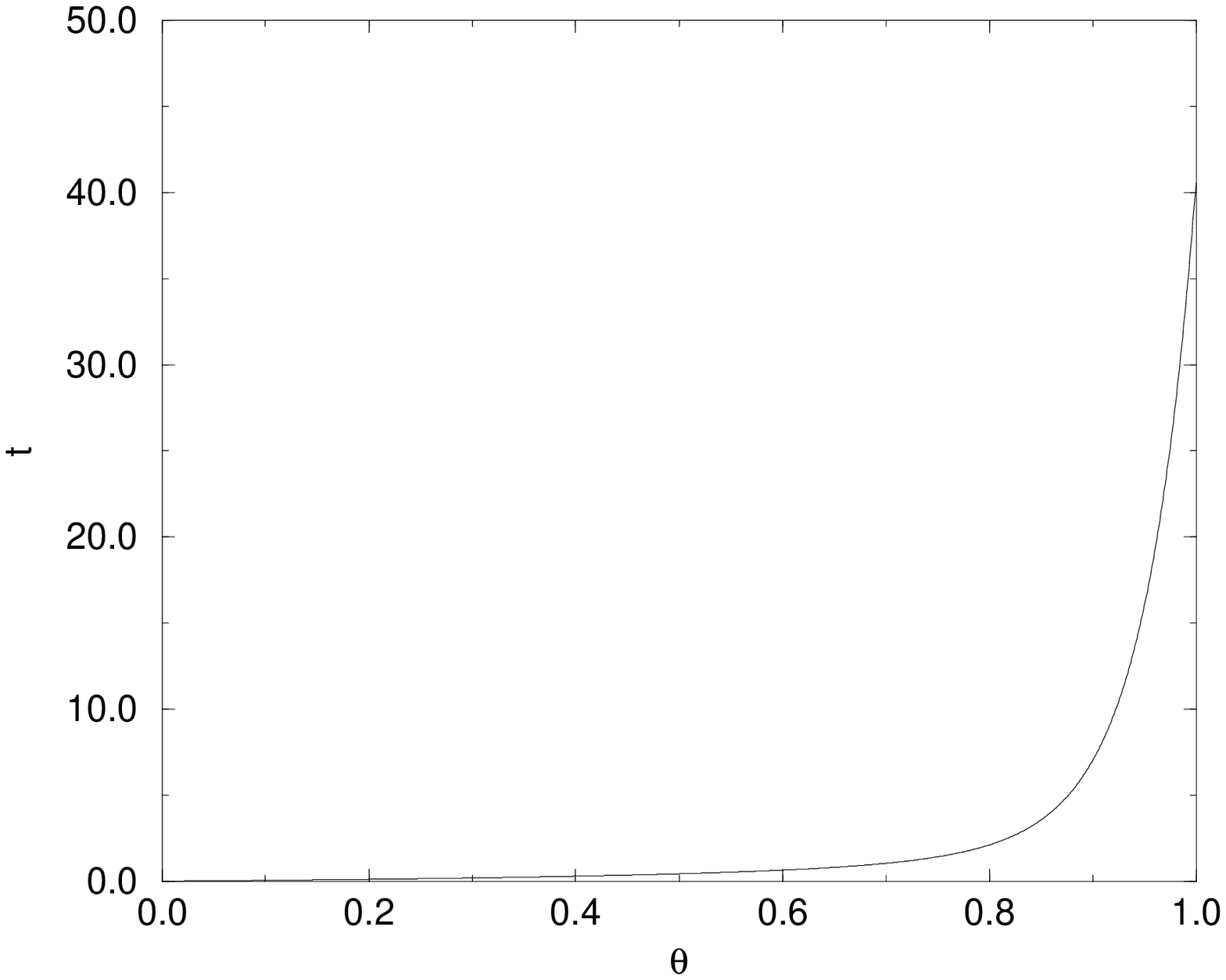,angle=90}
\end{center}
\end{minipage}
\end{figure}

\clearpage
\begin{figure}
\begin{minipage}[t]{16cm}
\begin{center}
FIG.3.
\vskip1cm
\epsfig{file=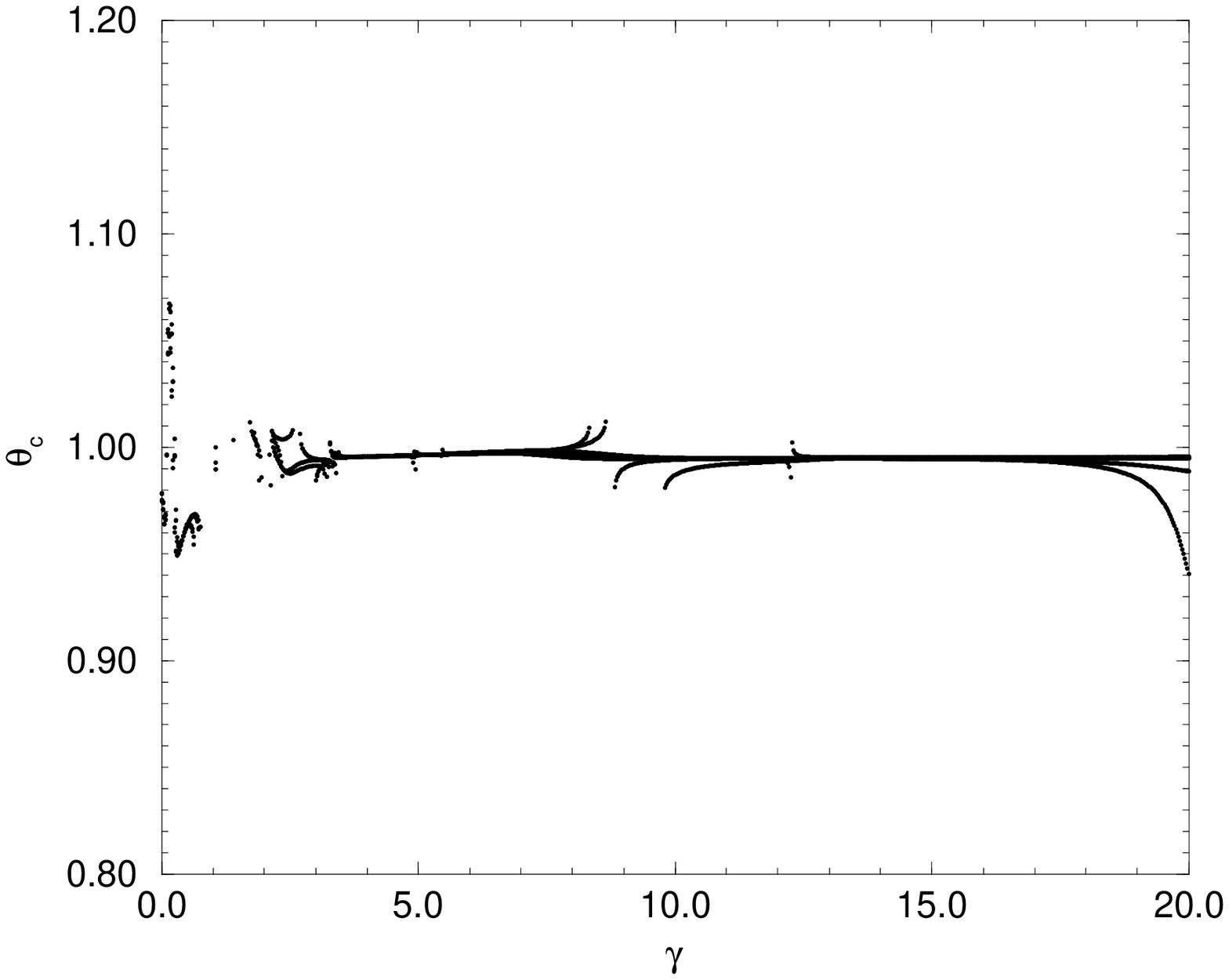,angle=90}
\end{center}
\end{minipage}
\end{figure}

\clearpage
\begin{figure}
\begin{minipage}[t]{16cm}
\begin{center}
FIG.4.
\vskip1cm
\epsfig{file=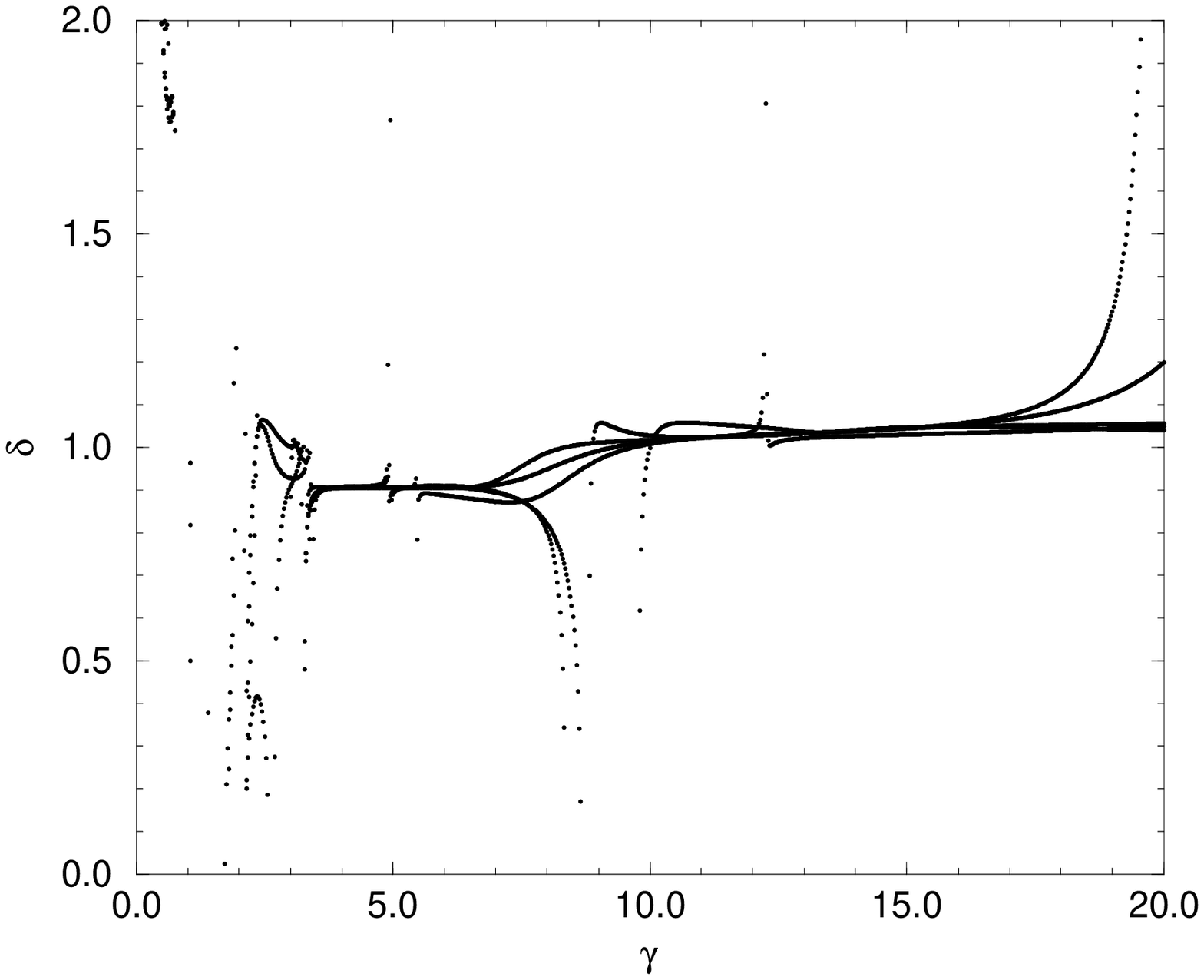,angle=90}
\end{center}
\end{minipage}
\end{figure}

\clearpage
\begin{figure}
\begin{minipage}[t]{16cm}
\begin{center}
FIG.5.
\vskip1cm
\epsfig{file=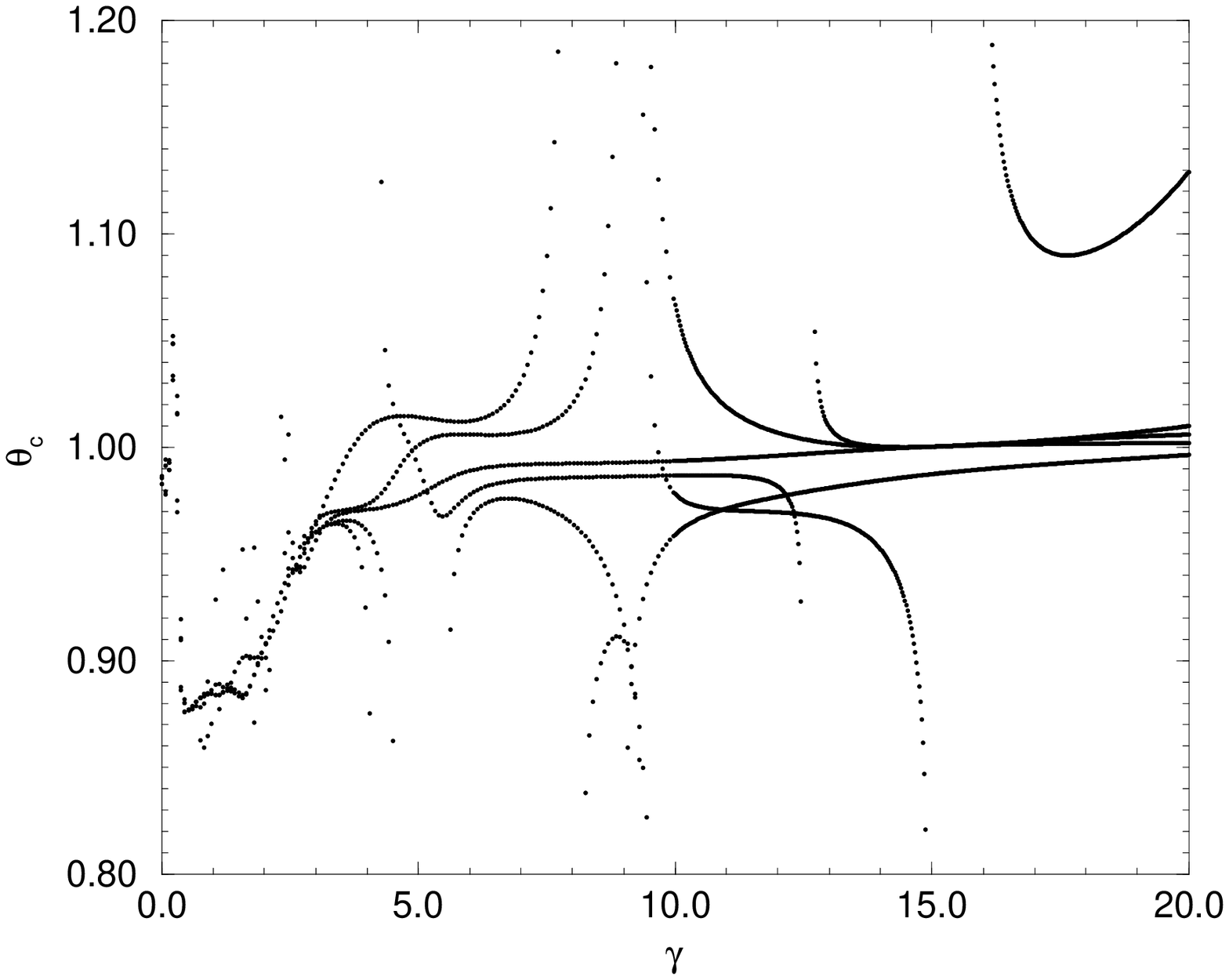,angle=90}
\end{center}
\end{minipage}
\end{figure}

\clearpage
\begin{figure}
\begin{minipage}[t]{16cm}
\begin{center}
FIG.6.
\vskip1cm
\epsfig{file=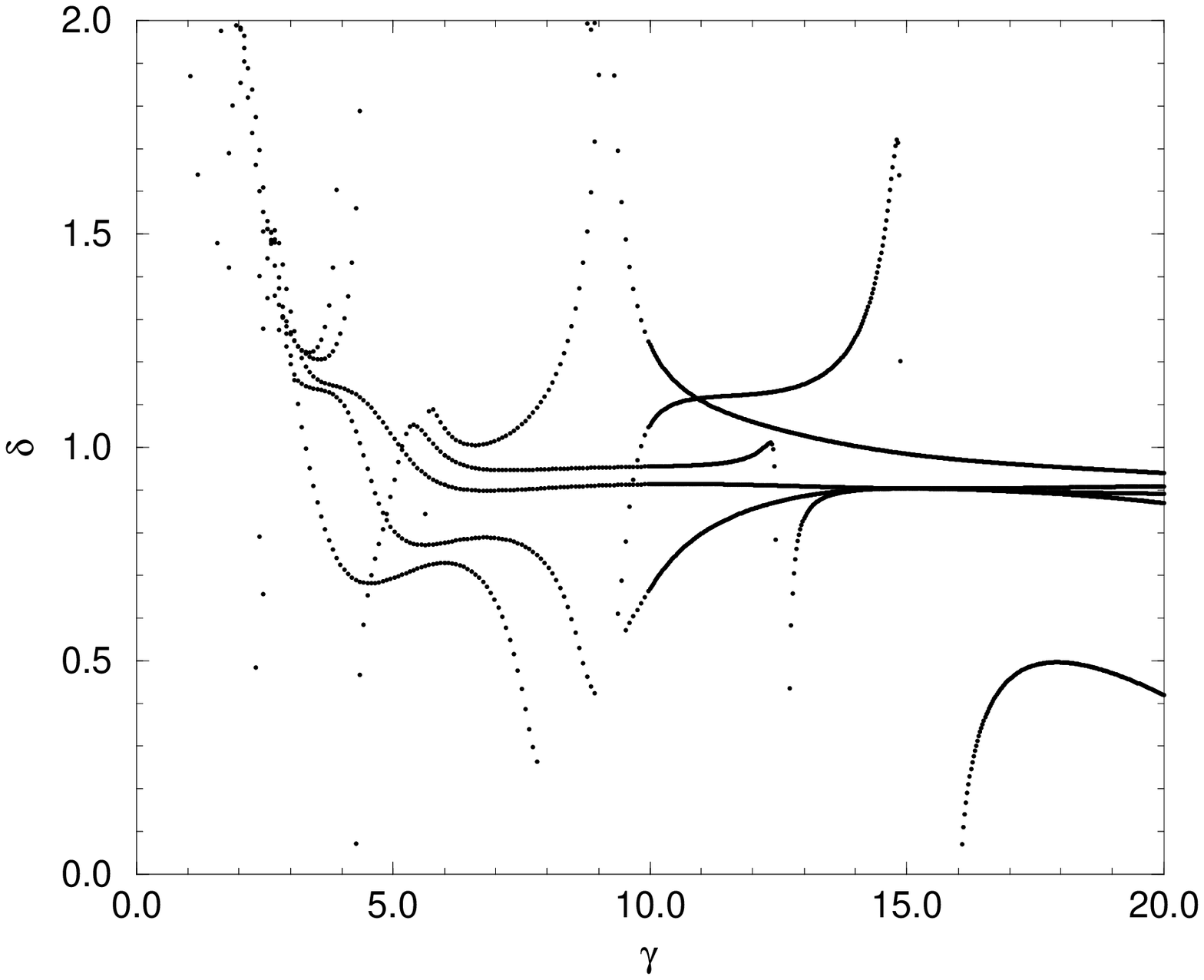,angle=90}
\end{center}
\end{minipage}
\end{figure}

\clearpage
\begin{figure}
\begin{minipage}[t]{16cm}
\begin{center}
FIG.7.
\vskip1cm
\epsfig{file=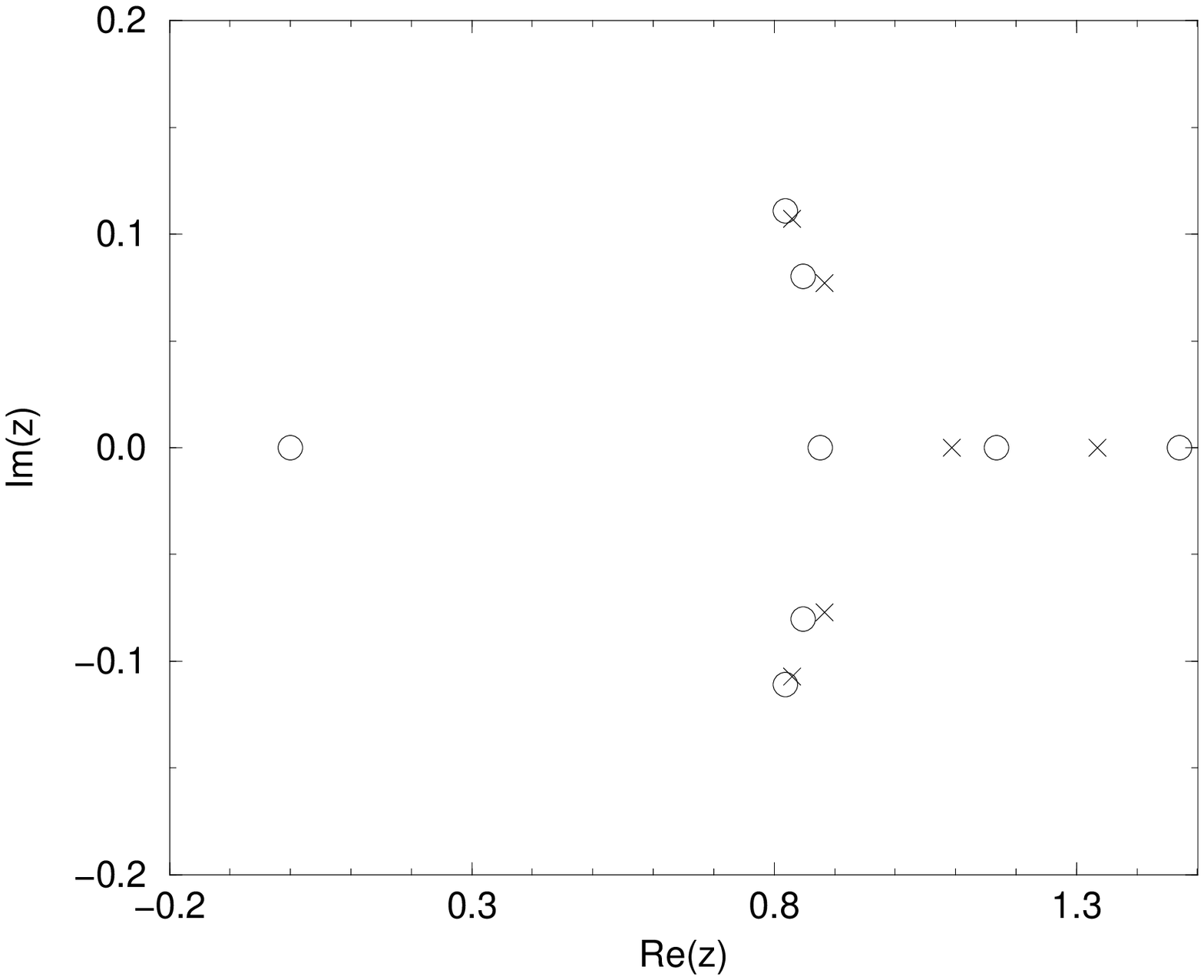,angle=90}
\end{center}
\end{minipage}
\end{figure}

\clearpage
\begin{figure}
\begin{minipage}[t]{16cm}
\begin{center}
FIG.8. 
\vskip1cm
\epsfig{file=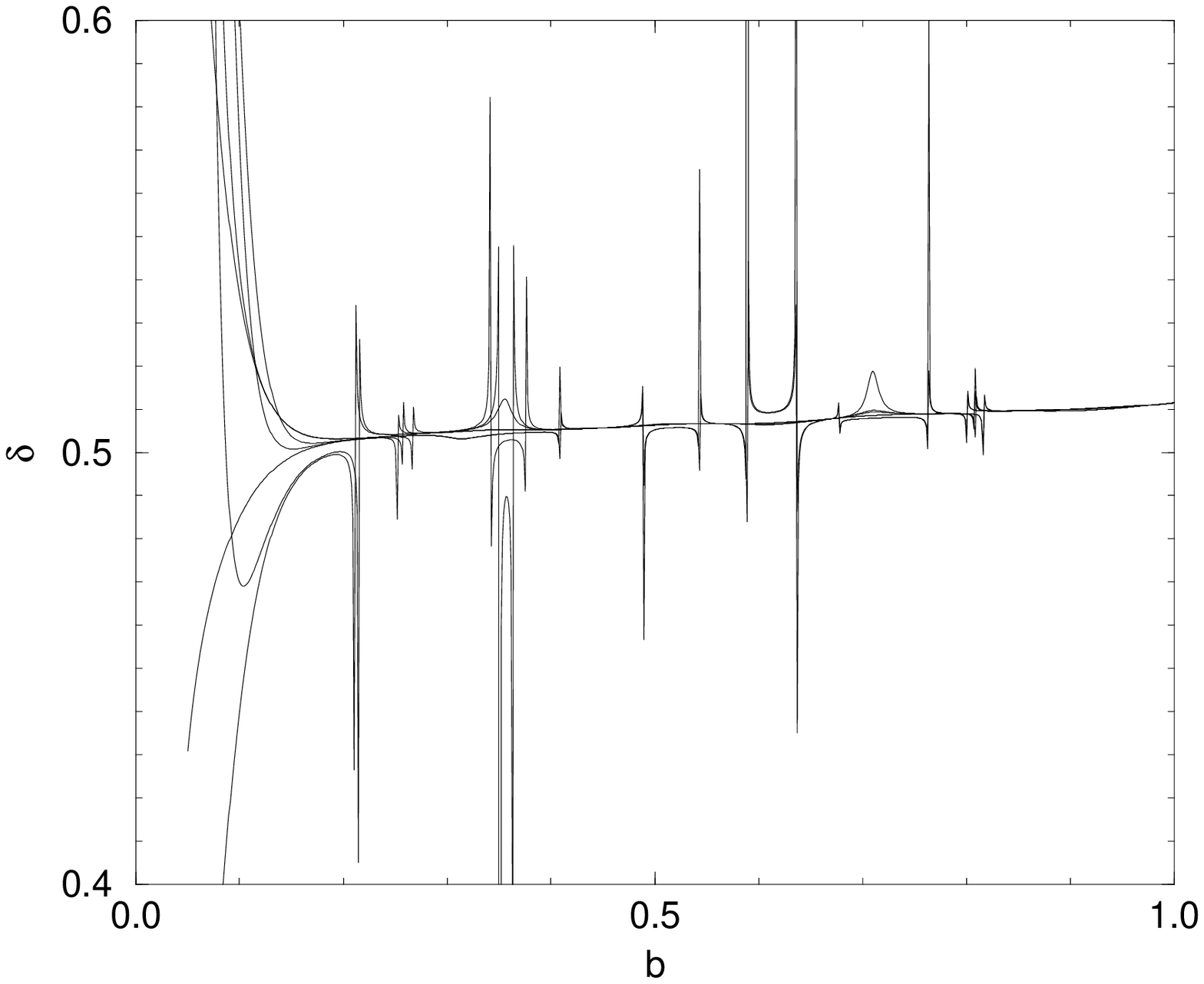,angle=90}
\end{center}
\end{minipage}
\end{figure}

\clearpage
\begin{figure}
\begin{minipage}[t]{16cm}
\begin{center}
FIG.9. 
\vskip1cm
\epsfig{file=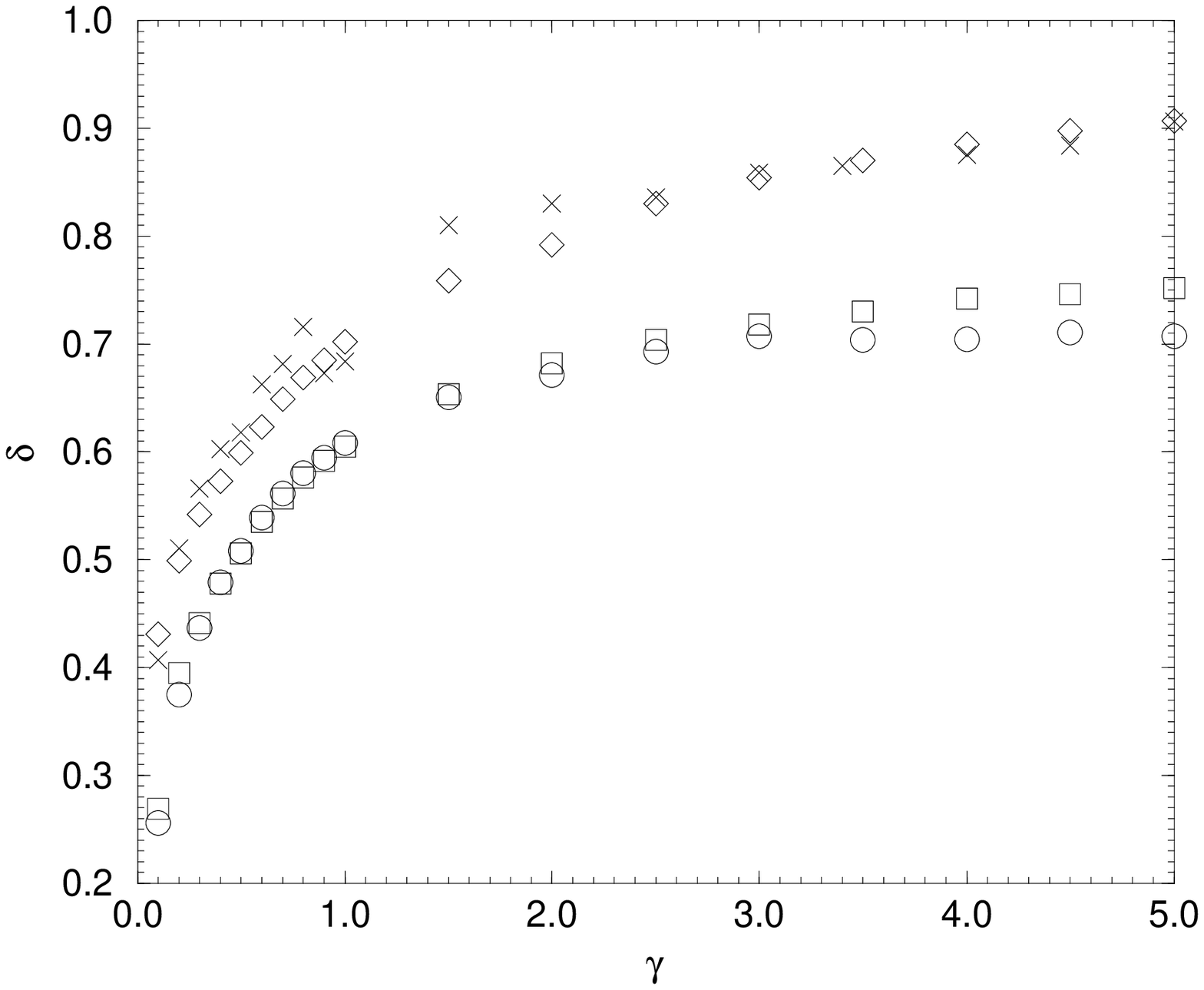,angle=90}
\end{center}
\end{minipage}
\end{figure}

\clearpage
\begin{figure}
\begin{minipage}[t]{16cm}
\begin{center}
FIG.10. 
\vskip1cm
\epsfig{file=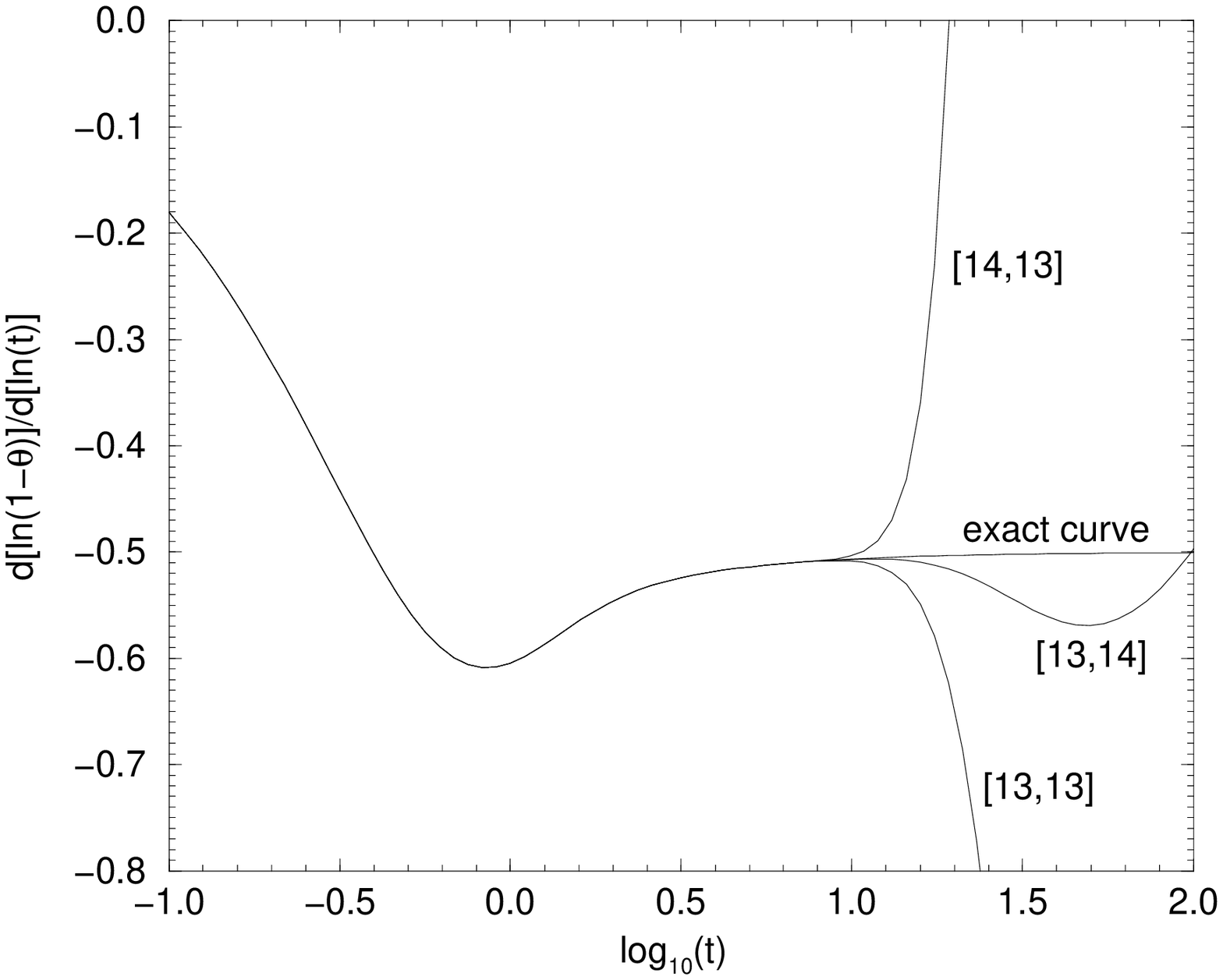,angle=90}
\end{center}
\end{minipage}
\end{figure}

\clearpage
\begin{figure}
\begin{minipage}[t]{16cm}
\begin{center}
FIG.11.
\vskip1cm
\epsfig{file=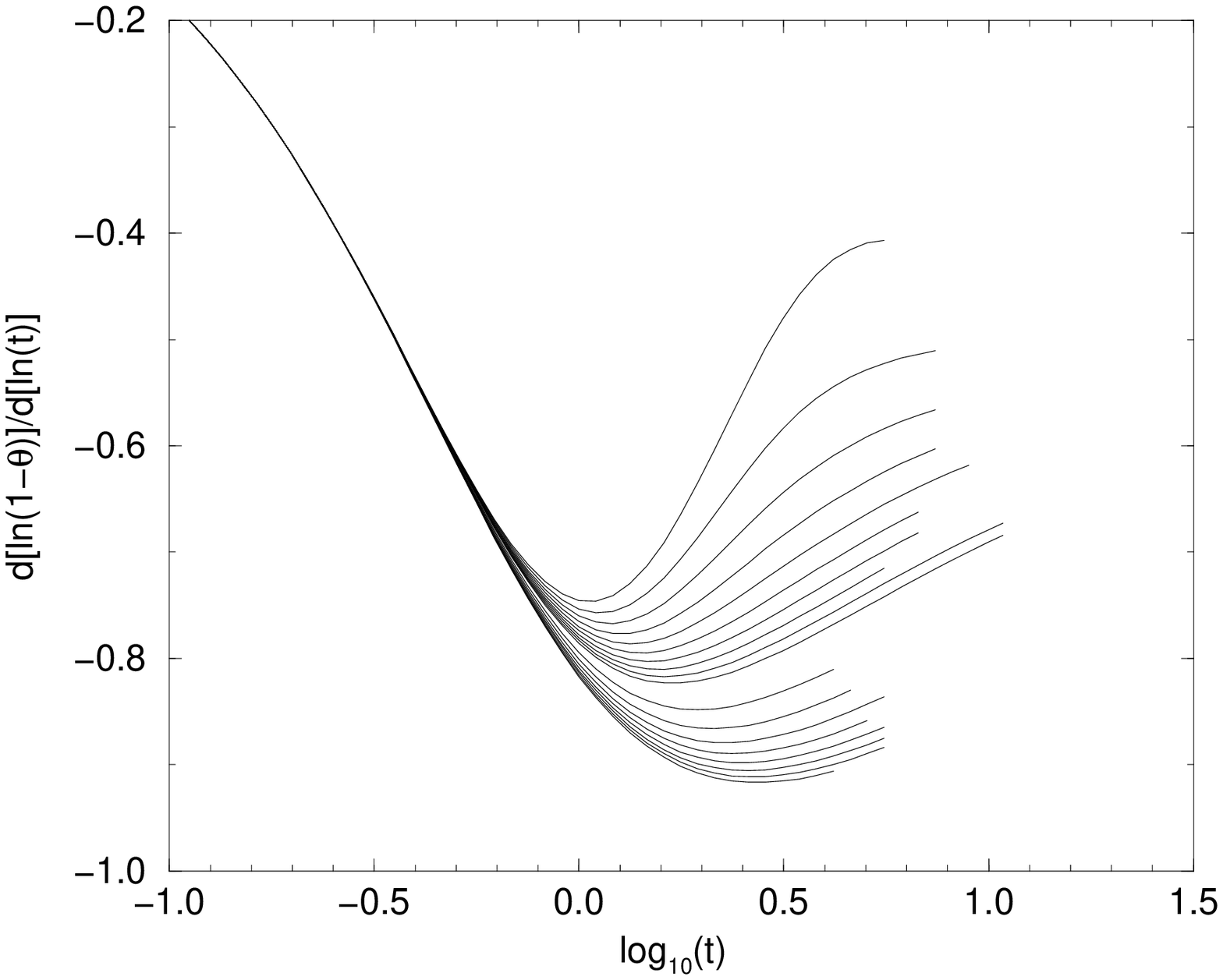,angle=90}
\end{center}
\end{minipage}
\end{figure}

\clearpage
\begin{figure}
\begin{minipage}[t]{16cm}
\begin{center}
FIG.12.
\vskip1cm
\epsfig{file=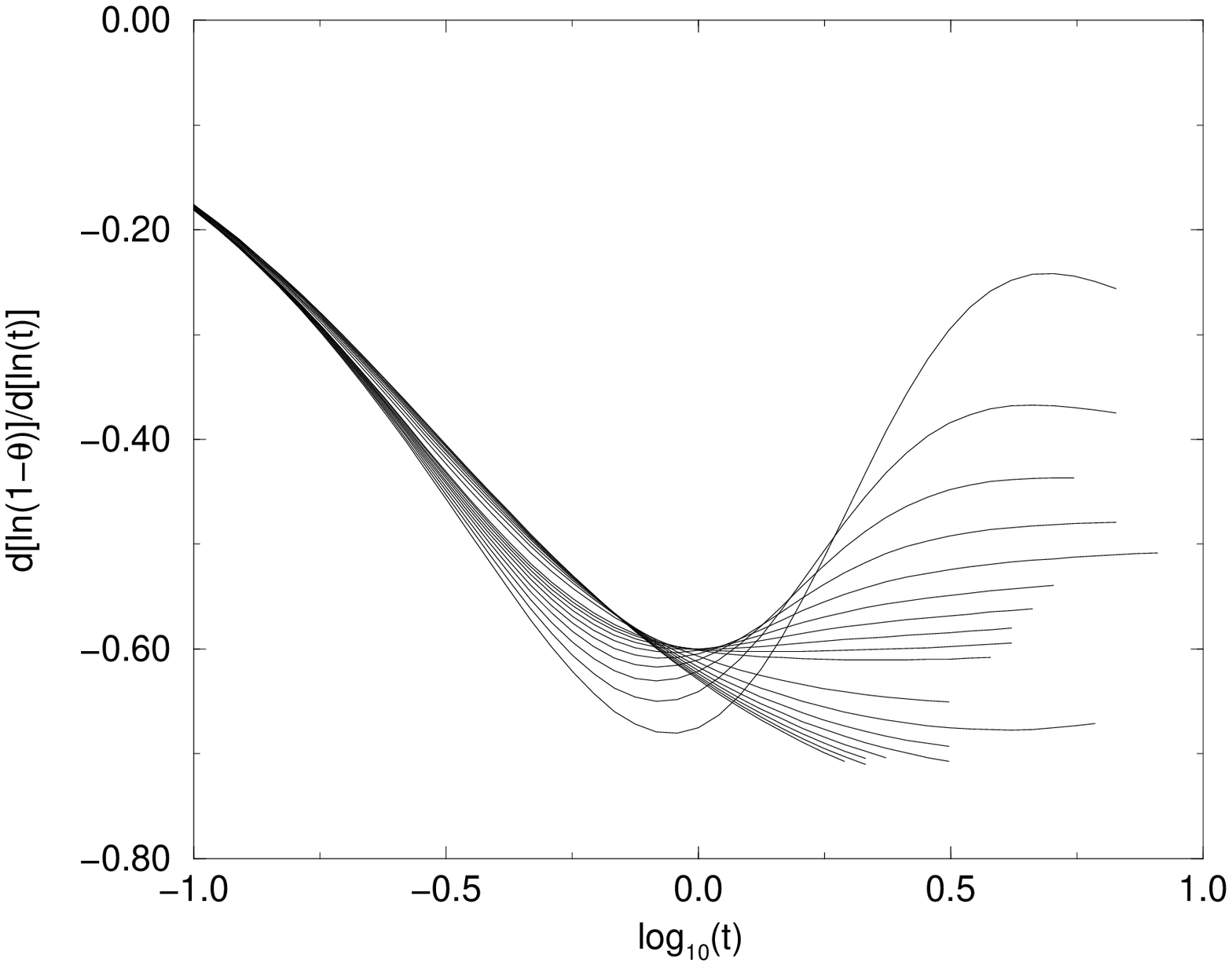,angle=90}
\end{center}
\end{minipage}
\end{figure}

\clearpage
\begin{figure}
\begin{minipage}[t]{16cm}
\begin{center}
FIG.13.
\vskip1cm
\epsfig{file=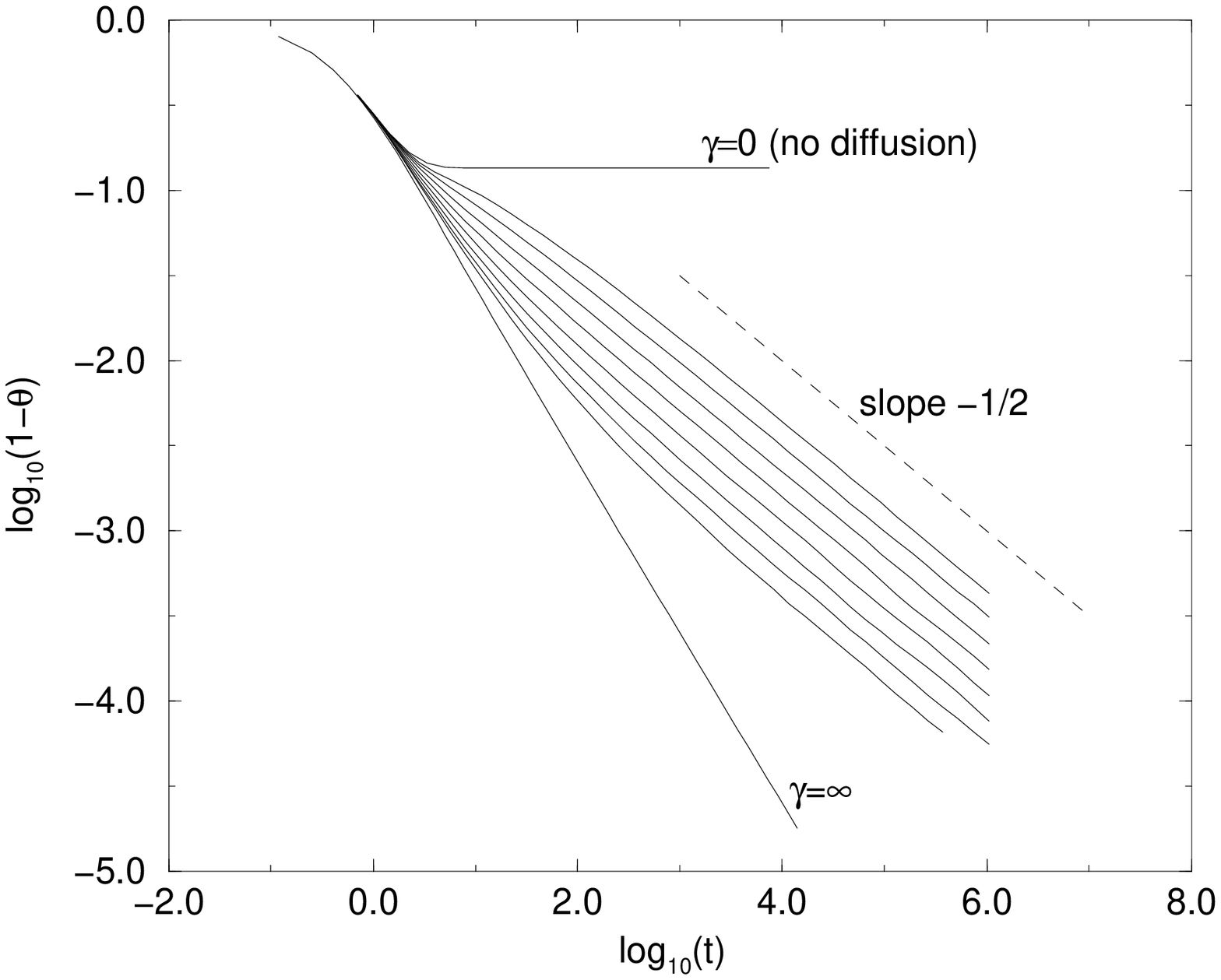,angle=90}
\end{center}
\end{minipage}
\end{figure}

\clearpage
\begin{figure}
\begin{minipage}[t]{16cm}
\begin{center}
FIG.14. 
\vskip1cm
\epsfig{file=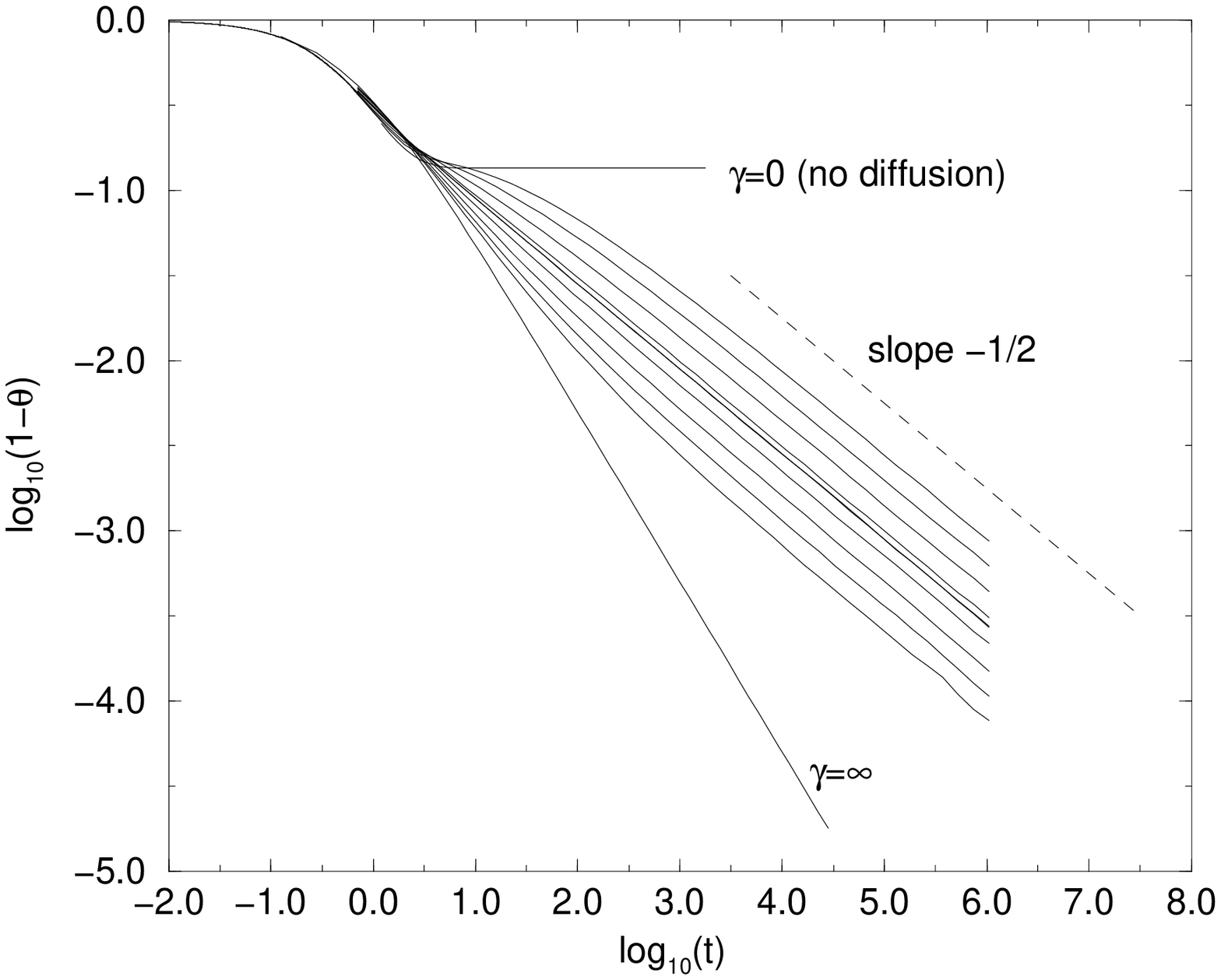,angle=90}
\end{center}
\end{minipage}
\end{figure}

\clearpage
\begin{figure}
\begin{minipage}[t]{16cm}
\begin{center}
FIG.15.
\vskip1cm
\epsfig{file=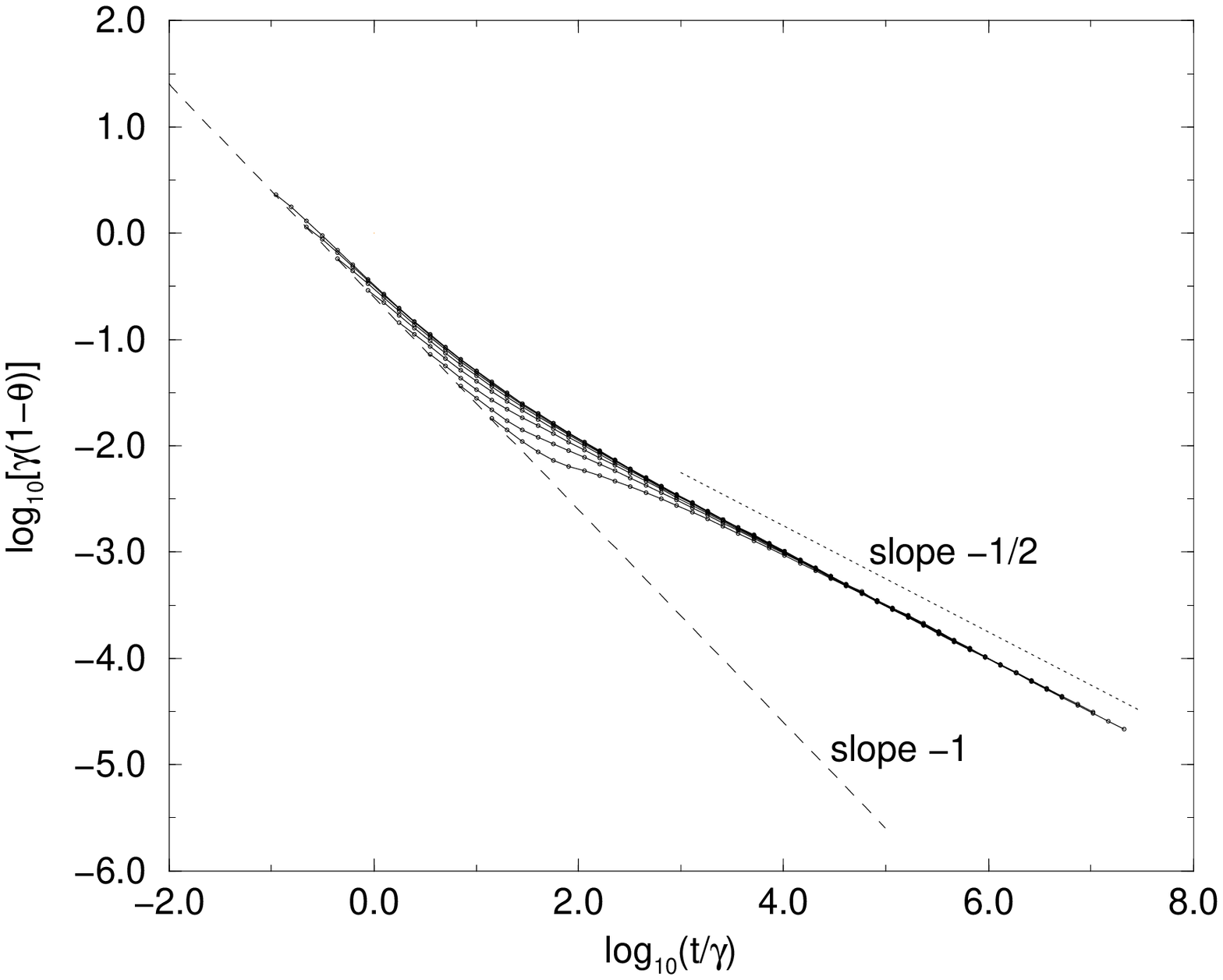,angle=90}
\end{center}
\end{minipage}
\end{figure}

\clearpage
\begin{figure}
\begin{minipage}[t]{16cm}
\begin{center}
FIG.16.
\vskip1cm
\epsfig{file=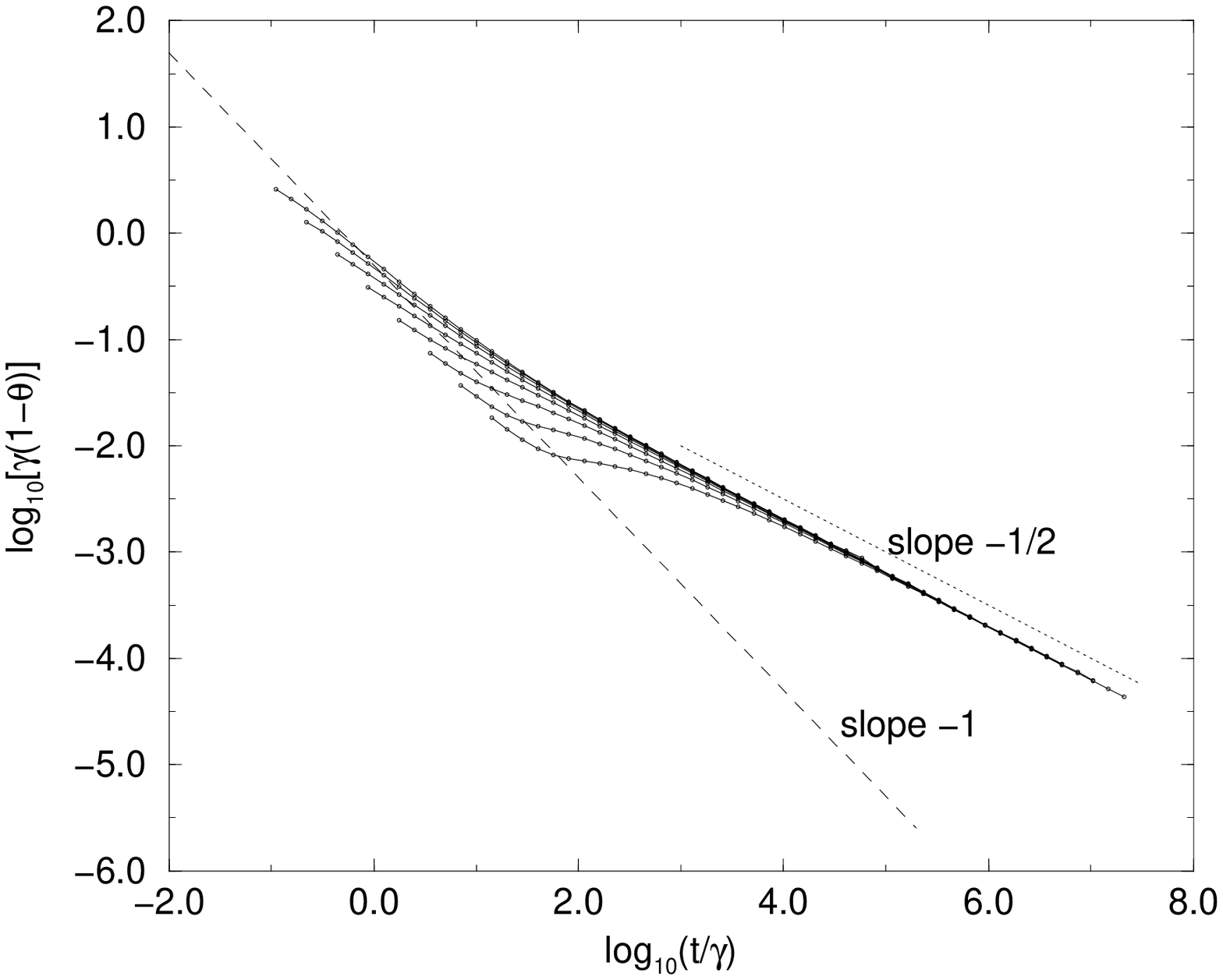,angle=90}
\end{center}
\end{minipage}
\end{figure}

\pagebreak
\clearpage
\appendix
\section{The series expansion coefficients for
diffusive dimer and monomer models}
\label{appendix:series_coefficient}
For the diffusive dimer model, the series expansion
coefficients $P(\o)^{(n)}$ for $n = 0, 1, 2, \ldots, 31$
are
\begin{eqnarray*}
{P(\o)}^{(0)} & = & 1,
\end{eqnarray*}
\begin{eqnarray*}
{P(\o)}^{(1)} & = & -2,
\end{eqnarray*}
\begin{eqnarray*}
{P(\o)}^{(2)} & = & 6,
\end{eqnarray*}
\begin{eqnarray*}
{P(\o)}^{(3)} & = & -22,
\end{eqnarray*}
\begin{eqnarray*}
{P(\o)}^{(4)} & = & 94,
\end{eqnarray*}
\begin{eqnarray*}
{P(\o)}^{(5)} & = & -454 - 8\gamma,
\end{eqnarray*}
\begin{eqnarray*}
{P(\o)}^{(6)} & = & 2430 + 136\gamma + 32\gamma^{2},
\end{eqnarray*}
\begin{eqnarray*}
{P(\o)}^{(7)} & = & -14214 - 1648\gamma - 544\gamma^{2} - 160\gamma^{3},
\end{eqnarray*}
\begin{eqnarray*}
{P(\o)}^{(8)} & = & 89918 + 17776\gamma + 6720\gamma^{2} + 2752\gamma^{3} + 896\gamma^{4},
\end{eqnarray*}
\begin{eqnarray*}
{P(\o)}^{(9)} & = & -610182 - 183640\gamma - 75488\gamma^{2} - 34848\gamma^{3} - 15808\gamma^{4}  \\
     &   &   - 5376\gamma^{5},
\end{eqnarray*}
\begin{eqnarray*}
{P(\o)}^{(10)} & = & 4412798 + 1876824\gamma + 829504\gamma^{2} + 406464\gamma^{3} + 207840\gamma^{4}  \\
      &   &     + 98560\gamma^{5} + 33792\gamma^{6},
\end{eqnarray*}
\begin{eqnarray*}
{P(\o)}^{(11)} & = & -33827974 - 19290560\gamma - 9193152\gamma^{2} - 4688896\gamma^{3}  \\
      &   &    - 2538496\gamma^{4} - 1364672\gamma^{5} - 651008\gamma^{6} - 219648\gamma^{7},
\end{eqnarray*}
\begin{eqnarray*}
{P(\o)}^{(12)} & = & 273646526 + 201202624\gamma + 104153760\gamma^{2} + 55017632\gamma^{3}  \\
      &   &    + 30829536\gamma^{4} + 17641472\gamma^{5} + 9613280\gamma^{6}  \\
      &   &    + 4487168\gamma^{7} + 1464320\gamma^{8},
\end{eqnarray*}
\begin{eqnarray*}
{P(\o)}^{(13)} & = & -2326980998 - 2140434856\gamma - 1212903168\gamma^{2}  \\
      &   &    - 664001632\gamma^{3} - 381899008\gamma^{4} - 226930976\gamma^{5}  \\
      &   &    - 132747296\gamma^{6} - 71354592\gamma^{7} - 31943680\gamma^{8} - 9957376\gamma^{9},
\end{eqnarray*}
\begin{eqnarray*}
{P(\o)}^{(14)} & = & 20732504062 + 23292327080\gamma + 14540797344\gamma^{2}  \\
      &   &    + 8274623072\gamma^{3} + 4870602688\gamma^{4} + 2973218272\gamma^{5}  \\
      &   &    + 1818085760\gamma^{6} + 1059060704\gamma^{7} + 550524320\gamma^{8}  \\
      &   &    + 233132032\gamma^{9} + 68796416\gamma^{10},
\end{eqnarray*}
\begin{eqnarray*}
{P(\o)}^{(15)} & = & -192982729350 - 259697613072\gamma - 179401122720\gamma^{2}  \\
      &   &    - 106542051936\gamma^{3} - 64143045632\gamma^{4} - 40012549568\gamma^{5}  \\
      &   &    - 25252420800\gamma^{6} - 15495356512\gamma^{7} - 8810633856\gamma^{8}  \\
      &   &    - 4369323616\gamma^{9} - 1734705152\gamma^{10} - 481574912\gamma^{11},
\end{eqnarray*}
\begin{eqnarray*}
{P(\o)}^{(16)} & = & 1871953992254 + 2969098816016\gamma + 2275429063808\gamma^{2}  \\
      &   &    + 1416343409184\gamma^{3} + 872645113632\gamma^{4} + 554654077056\gamma^{5}  \\
      &   &    + 358789726592\gamma^{6} + 228594111904\gamma^{7} + 137952689120\gamma^{8}  \\
      &   &    + 75460620608\gamma^{9} + 35391538304\gamma^{10} + 13105004544\gamma^{11}  \\
      &   &    + 3408068608\gamma^{12},
\end{eqnarray*}
\begin{eqnarray*}
{P(\o)}^{(17)} & = & -18880288847750 - 34819585889272\gamma - 29629807780320\gamma^{2}  \\
      &   &    - 19414284425280\gamma^{3} - 12257502997152\gamma^{4} - 7925353240960\gamma^{5}  \\
      &   &    - 5232190291296\gamma^{6} - 3433400249408\gamma^{7} - 2164670388352\gamma^{8}  \\
      &   &    - 1265163225632\gamma^{9} - 659040825216\gamma^{10} - 290843555840\gamma^{11}  \\
      &   &    - 100193632256\gamma^{12} - 24343347200\gamma^{13},
\end{eqnarray*}
\begin{eqnarray*}
{P(\o)}^{(18)} & = & 197601208474238 + 418857922740216\gamma + 395602299173696\gamma^{2}  \\
      &   &    + 273978158172160\gamma^{3} + 177588283397088\gamma^{4}  \\
      &   &    + 116718001713792\gamma^{5} + 78414883481952\gamma^{6}  \\
      &   &    + 52726551390048\gamma^{7} + 34420850096512\gamma^{8} + 21138424854304\gamma^{9}  \\
      &   &    + 11830865489344\gamma^{10} + 5828807502880\gamma^{11} + 2414315305984\gamma^{12}  \\
      &   &    + 773310775296\gamma^{13} + 175272099840\gamma^{14},
\end{eqnarray*}
\begin{eqnarray*}
{P(\o)}^{(19)} & = & -2142184050841734 - 5167334116337248\gamma  \\
      &   &    - 5409353469693312\gamma^{2} - 3974358277418464\gamma^{3}  \\
      &   &    - 2650710992073376\gamma^{4} - 1770760359697088\gamma^{5}  \\
      &   &    - 1208250123782816\gamma^{6} - 829498459619104\gamma^{7}  \\
      &   &    - 557514170611008\gamma^{8} - 356342440781248\gamma^{9}  \\
      &   &    - 210643968436256\gamma^{10} - 111995901788096\gamma^{11}  \\
      &   &    - 51951090626080\gamma^{12} - 20179683365760\gamma^{13}  \\
      &   &    - 6013648437248\gamma^{14} - 1270722723840\gamma^{15},
\end{eqnarray*}
\begin{eqnarray*}
{P(\o)}^{(20)} & = & 24016181943732414 + 65354875319253600\gamma + 75674015695071776\gamma^{2}  \\
      &   &    + 59170426635099072\gamma^{3} + 40709461342679712\gamma^{4}  \\
      &   &    + 27654553375343168\gamma^{5} + 19139165182573632\gamma^{6}  \\
      &   &    + 13379477948777856\gamma^{7} + 9220546462388512\gamma^{8}  \\
      &   &    + 6096023066059456\gamma^{9} + 3768570926043360\gamma^{10}  \\
      &   &    + 2126161230748448\gamma^{11} + 1067953378669088\gamma^{12}  \\
      &   &    + 465021052820384\gamma^{13} + 169438907614560\gamma^{14}  \\
      &   &    + 47047244775424\gamma^{15} + 9268801044480\gamma^{16},
\end{eqnarray*}
\begin{eqnarray*}
{P(\o)}^{(21)} & = & -278028611833689478 - 847070521919796296\gamma  \\
      &   &   - 1082140708741735168\gamma^{2} - 902807166330596704\gamma^{3}  \\
      &   &   - 642463112877878368\gamma^{4} - 444202148035295936\gamma^{5}  \\
      &   &   - 311557082954071776\gamma^{6} - 221310190035972736\gamma^{7}  \\
      &   &   - 155891975926137920\gamma^{8} - 106130622734975296\gamma^{9}  \\
      &   &   - 68163212990088160\gamma^{10} - 40401083210384096\gamma^{11}  \\
      &   &   - 21628549879148512\gamma^{12} - 10224685696979776\gamma^{13}  \\
      &   &   - 4170519535181536\gamma^{14} - 1426800665301088\gamma^{15}  \\
      &   &   - 369849931661312\gamma^{16} - 67971207659520\gamma^{17},
\end{eqnarray*}
\begin{eqnarray*}
{P(\o)}^{(22)} & = & 3319156078802044158 + 11245724095683198280\gamma  \\
      &   &    + 15806297066594859744\gamma^{2} + 14097805354882121664\gamma^{3}  \\
      &   &    + 10405357836443942528\gamma^{4} + 7331412297306645056\gamma^{5}  \\
      &   &    + 5209126487226619872\gamma^{6} + 3753693513361866688\gamma^{7}  \\
      &   &    + 2695682775965810496\gamma^{8} + 1883121553855520800\gamma^{9}  \\
      &   &    + 1250352187435821024\gamma^{10} + 773119232263912256\gamma^{11}  \\
      &   &    + 436759047669796704\gamma^{12} + 221031829068739616\gamma^{13}  \\
      &   &    + 98090146080366016\gamma^{14} + 37415367315785664\gamma^{15}  \\
      &   &    + 12034887222344992\gamma^{16} + 2918797259309056\gamma^{17}  \\
      &   &    + 500840477491200\gamma^{18},
\end{eqnarray*}
\begin{eqnarray*}
{P(\o)}^{(23)} & = & -40811417293301014150 - 152849819143271186864\gamma  \\
      &   &    - 235669362697015641504\gamma^{2} - 225030434256866474048\gamma^{3}  \\
      &   &    - 172730720539355533536\gamma^{4} - 124209020387171547840\gamma^{5}  \\
      &   &    - 89396791275859021472\gamma^{6} - 65265468008645791072\gamma^{7}  \\
      &   &    - 47681226234984815648\gamma^{8} - 34078914479620848544\gamma^{9}  \\
      &   &    - 23302022034195791264\gamma^{10} - 14950154580712363008\gamma^{11}  \\
      &   &    - 8846189597178252768\gamma^{12} - 4745848875738292928\gamma^{13}  \\
      &   &    - 2265231509811056512\gamma^{14} - 941951799608143040\gamma^{15}  \\
      &   &    - 335416253872860704\gamma^{16} - 101595547064760928\gamma^{17}  \\
      &   &    - 23107112699691008\gamma^{18} - 3706219533434880\gamma^{19},
\end{eqnarray*}
\begin{eqnarray*}
{P(\o)}^{(24)} & = & 516247012345341914942 + 2125833345702860926128\gamma  \\
      &   &    + 3584709736316916197120\gamma^{2} + 3667635825937886265248\gamma^{3}  \\
      &   &    + 2935302250997270008320\gamma^{4} + 2157911658548001672192\gamma^{5}  \\
      &   &    + 1573612047377468555296\gamma^{6} + 1162780482970536810112\gamma^{7}  \\
      &   &    + 862635165194788176384\gamma^{8} + 629254138123646209984\gamma^{9}  \\
      &   &    + 441670494784119415040\gamma^{10} + 292773697423309629888\gamma^{11}  \\
      &   &    + 180386574048628422176\gamma^{12} + 101774434388607688384\gamma^{13}  \\
      &   &    + 51747976608776753696\gamma^{14} + 23266775394437326144\gamma^{15}  \\
      &   &    + 9052145344807326144\gamma^{16} + 3002516812188788608\gamma^{17}  \\
      &   &    + 857815531941631296\gamma^{18} + 183397309348839424\gamma^{19}  \\
      &   &    + 27531916534087680\gamma^{20},
\end{eqnarray*}
\begin{eqnarray*}
{P(\o)}^{(25)} & = & -6711185258405244576646 - 30238180002704596333208\gamma  \\
      &   &    - 55598200570861707979680\gamma^{2} - 60976389483506749681568\gamma^{3}  \\
      &   &    - 51003312645322864268128\gamma^{4} - 38404740494773062950176\gamma^{5}  \\
      &   &    - 28389964187770573475040\gamma^{6} - 21217507587703139716128\gamma^{7}  \\
      &   &    - 15959639632228805304256\gamma^{8} - 11856781604498555126144\gamma^{9}  \\
      &   &    - 8519923504665486348256\gamma^{10} - 5814814052946982013216\gamma^{11}  \\
      &   &    - 3713060409585399321472\gamma^{12} - 2188983105017137174464\gamma^{13}  \\
      &   &    - 1175478617129581490560\gamma^{14} - 566022591719541854336\gamma^{15}  \\
      &   &    - 239600608251046218528\gamma^{16} - 87093953947019843424\gamma^{17}  \\
      &   &    - 26826301792524728832\gamma^{18} - 7241192306564087872\gamma^{19}  \\
      &   &    - 1458610850414723072\gamma^{20} - 205237923254108160\gamma^{21},
\end{eqnarray*}
\begin{eqnarray*}
{P(\o)}^{(26)} & = & 89574471680939133937534 + 439663978304010761309336\gamma  \\
      &   &    + 878864969592056661663680\gamma^{2} + 1033220218894344066221152\gamma^{3}  \\
      &   &    + 905162593426518959688480\gamma^{4} + 699466982856423681626592\gamma^{5}  \\
      &   &    + 524550622363351486026048\gamma^{6} + 396321792642538988104928\gamma^{7}  \\
      &   &    + 301868194425498598076992\gamma^{8} + 227989242191471816249696\gamma^{9}  \\
      &   &    + 167333371520455599867424\gamma^{10} + 117243269450611152416000\gamma^{11}  \\
      &   &    + 77295857780714519060928\gamma^{12} + 47370056124944928474784\gamma^{13}  \\
      &   &    + 26675453205517382525856\gamma^{14} + 13627834185309074754720\gamma^{15}  \\
      &   &    + 6215448113411854023776\gamma^{16} + 2476949102568415569728\gamma^{17}  \\
      &   &    + 839899058850812411936\gamma^{18} + 239165277641087918144\gamma^{19}  \\
      &   &    + 61092781558844261824\gamma^{20} + 11620352278518562816\gamma^{21}  \\
      &   &    + 1534822730422026240\gamma^{22},
\end{eqnarray*}
\begin{eqnarray*}
{P(\o)}^{(27)} & = & -1226366187219563392423046 - 6531379936679608555490048\gamma  \\
      &   &    - 14153158560229748831123712\gamma^{2} - 17829942516341618875683488\gamma^{3}  \\
      &   &    - 16390388657713777384337248\gamma^{4} - 13024188161033852626760832\gamma^{5}  \\
      &   &    - 9918027745419496427963072\gamma^{6} - 7573935646469621224533504\gamma^{7}  \\
      &   &    - 5835396744531198025370240\gamma^{8} - 4473355152862831429807584\gamma^{9}  \\
      &   &    - 3346869880800982736546432\gamma^{10} - 2401496033284393106394304\gamma^{11}  \\
      &   &    - 1629516516241948614285760\gamma^{12} - 1033801940311680754546112\gamma^{13}  \\
      &   &    - 607038025256747982659264\gamma^{14} - 326452128844455325407648\gamma^{15}  \\
      &   &    - 158757402409409198571968\gamma^{16} - 68639251401951128448448\gamma^{17}  \\
      &   &    - 25759744940242270466976\gamma^{18} - 8133338764734443700064\gamma^{19}  \\
      &   &    - 2127377128471631380160\gamma^{20} - 515040012051367812096\gamma^{21}  \\
      &   &    - 92704039977088974848\gamma^{22} - 11511170478165196800\gamma^{23},
\end{eqnarray*}
\begin{eqnarray*}
{P(\o)}^{(28)} & = & 17208434165059531880467902 + 99081506420313710986772736\gamma  \\
      &   &    + 232103903429219081532907872\gamma^{2} + 313143052910337271943545632\gamma^{3}  \\
      &   &    + 302531882137452591255077248\gamma^{4} + 247695203872064524469329920\gamma^{5}  \\
      &   &    + 191751044433982602086189184\gamma^{6} + 148003121016823516846598336\gamma^{7}  \\
      &   &    + 115246426164717743454533376\gamma^{8} + 89548623260586320065607520\gamma^{9}  \\
      &   &    + 68178700940595222195124032\gamma^{10} + 49993383490006417316624128\gamma^{11}  \\
      &   &    + 34821933734970293328723264\gamma^{12} + 22791830094404699697241952\gamma^{13}  \\
      &   &    + 13891323265031765512465344\gamma^{14} + 7814665824226532852984960\gamma^{15}  \\
      &   &    + 4016982900101880080201440\gamma^{16} + 1862205866467120855344800\gamma^{17}  \\
      &   &    + 764317776739694865201344\gamma^{18} + 270291929328683129059200\gamma^{19}  \\
      &   &    + 79296629516951459720032\gamma^{20} + 18879546837921695785504\gamma^{21}  \\
      &   &    + 4338110967536077063424\gamma^{22} + 740404461632374177792\gamma^{23}  \\
      &   &    + 86564001995802279936\gamma^{24},
\end{eqnarray*}
\begin{eqnarray*}
{P(\o)}^{(29)} & = & -247289888972538586949878150 - 1534172692240041452626636520\gamma  \\
      &   &    - 3874787071735125029536527360\gamma^{2} - 5593914936389904746992944000\gamma^{3}  \\
      &   &    - 5687097195642308827551818432\gamma^{4} - 4806905879064545608139517504\gamma^{5}  \\
      &   &    - 3787785163974288880030601664\gamma^{6} - 2955603108243908695631102784\gamma^{7}  \\
      &   &    - 2324464815975057231796395232\gamma^{8} - 1828539431234039817220631104\gamma^{9}  \\
      &   &    - 1414541083036286620968100704\gamma^{10} - 1058040639042272841296197632\gamma^{11}  \\
      &   &    - 754795598374307863822508064\gamma^{12} - 508234645790343970597871232\gamma^{13}  \\
      &   &    - 320323245811479831435568800\gamma^{14} - 187552827879867728682574240\gamma^{15}  \\
      &   &    - 101194015894779754258916928\gamma^{16} - 49808752256243832593146240\gamma^{17}  \\
      &   &    - 22057128789803220776595968\gamma^{18} - 8610553430581355224855168\gamma^{19}  \\
      &   &    - 2871992002592102396999904\gamma^{20} - 781079045618282247374272\gamma^{21}  \\
      &   &    - 167173262847020276611008\gamma^{22} - 36502951600975155899648\gamma^{23}  \\
      &   &    - 5918914576756420640768\gamma^{24} - 652559399660663341056\gamma^{25},
\end{eqnarray*}
\begin{eqnarray*}
{P(\o)}^{(30)} & = & 3636599975026505414628377086 + 24235192834488592555063381480\gamma  \\
      &   &    + 65825866944987465106280540640\gamma^{2} + 101588178322920631379346693376\gamma^{3}  \\
      &   &    + 108792583977960919823164346304\gamma^{4} + 95106347438122321463161286528\gamma^{5}  \\
      &   &    + 76389237086435129240476170944\gamma^{6} + 60282902962582074896568152544\gamma^{7}  \\
      &   &    + 47861156428195317146948964544\gamma^{8} + 38077374671506940952543821152\gamma^{9}  \\
      &   &    + 29889064803981070602563425824\gamma^{10} + 22767753081638239927853052832\gamma^{11}  \\
      &   &    + 16603099771615247450036647104\gamma^{12} + 11473120418073147648994092160\gamma^{13}  \\
      &   &    + 7454372028423481067708753792\gamma^{14} + 4523940784361767910490004224\gamma^{15}  \\
      &   &    + 2547674890838510347194642304\gamma^{16} + 1321026477071424198196864544\gamma^{17}  \\
      &   &    + 624158962537014248104138272\gamma^{18} + 264717889700048519760895712\gamma^{19}  \\
      &   &    + 98517274209783711927693664\gamma^{20} + 31032886905928789132563200\gamma^{21}  \\
      &   &    + 7806681142471034370768032\gamma^{22} + 1477165355656235095757792\gamma^{23}  \\
      &   &    + 306829936381452586376960\gamma^{24} + 47352796340581207900160\gamma^{25} \\
      &   &    + 4930448797436123021312\gamma^{26},
\end{eqnarray*}
\begin{eqnarray*}
{P(\o)}^{(31)} & = & -54690132113431117456486546054 - 390402306358974047554407998032\gamma  \\
      &   &    - 1137582424553489289447807997792\gamma^{2}  - 1874665509618062031115241291296\gamma^{3}  \\
      &   &    - 2116316617323806096783903601344\gamma^{4}  - 1916826452624015867653776105344\gamma^{5}  \\
      &   &    - 1571619063016185146258668134144\gamma^{6} - 1255045527053951534543760304384\gamma^{7}  \\
      &   &    - 1005606717413244521960545134816\gamma^{8} - 808411234967533218962071194176\gamma^{9}  \\
      &   &    - 643116053649141335307306261344\gamma^{10} - 498189822261377252029700331328\gamma^{11}  \\
      &   &    - 370737701124192040949130562656\gamma^{12} - 262369454324949422539656437184\gamma^{13}  \\
      &   &    - 175268684516776147755547530080\gamma^{14} - 109871960830604539165436865664\gamma^{15}  \\
      &   &    - 64284643706828638935486335392\gamma^{16} - 34893515783410832792353344544\gamma^{17}  \\
      &   &    - 17434380523486099607832839808\gamma^{18} - 7930481323762069309980955744\gamma^{19}  \\
      &   &    - 3230836698378503824136216160\gamma^{20} - 1149328018944519338194160160\gamma^{21}  \\
      &   &    - 342504050878779281408793504\gamma^{22} - 79568648016407862847405792\gamma^{23}  \\
      &   &    - 13027718319821439160157152\gamma^{24} - 2576298242354693706469632\gamma^{25}  \\
      &   &    - 379070964781663249235968\gamma^{26}  - 37330540894873502875648\gamma^{27}.
\end{eqnarray*}
For the diffusive monomer model, we have 
\begin{eqnarray*}
{P(\o)}^{(0)} & = & 1,
\end{eqnarray*}
\begin{eqnarray*}
{P(\o)}^{(1)} & = & -2, 
\end{eqnarray*}
\begin{eqnarray*}
{P(\o)}^{(2)} & = & 6,
\end{eqnarray*}
\begin{eqnarray*}
{P(\o)}^{(3)} & = & -22 + 4\gamma,
\end{eqnarray*}
\begin{eqnarray*}
{P(\o)}^{(4)} & = & 94 - 40\gamma - 16\gamma^{2},
\end{eqnarray*}
\begin{eqnarray*}
{P(\o)}^{(5)} & = & -454 + 316\gamma + 136\gamma^{2} + 80\gamma^{3},
\end{eqnarray*}
\begin{eqnarray*}
{P(\o)}^{(6)} & = & 2430 - 2384\gamma - 840\gamma^{2} - 608\gamma^{3} - 448\gamma^{4},
\end{eqnarray*}
\begin{eqnarray*}
{P(\o)}^{(7)} & = & -14214 + 18116\gamma + 4240\gamma^{2} + 3072\gamma^{3} + 3136\gamma^{4} + 2688\gamma^{5},
\end{eqnarray*}
\begin{eqnarray*}
{P(\o)}^{(8)} & = & 89918 - 141432\gamma - 13920\gamma^{2} - 9728\gamma^{3} - 13120\gamma^{4} - 17664\gamma^{5} - 16896\gamma^{6},
\end{eqnarray*}
\begin{eqnarray*}
{P(\o)}^{(9)} & = & -610182 + 1143564\gamma - 52040\gamma^{2} - 21072\gamma^{3} 
+ 16416\gamma^{4} + 60480\gamma^{5} \\
     &   & + 105600\gamma^{6} + 109824\gamma^{7},
\end{eqnarray*}
\begin{eqnarray*}
{P(\o)}^{(10)} & = & 4412798 - 9606304\gamma + 1860776\gamma^{2} + 776864\gamma^{3}  \\
      &   &  + 366240\gamma^{4} + 76544\gamma^{5} - 285824\gamma^{6} - 658944\gamma^{7} - 732160\gamma^{8},
\end{eqnarray*}
\begin{eqnarray*}
{P(\o)}^{(11)} & = & -33827974 + 83906644\gamma - 28877184\gamma^{2} - 9589600\gamma^{3} \\
      &   & - 5139808\gamma^{4} - 3368896\gamma^{5} - 1464064\gamma^{6}  \\
      &   &  + 1317888\gamma^{7} + 4246528\gamma^{8} + 4978688\gamma^{9},
\end{eqnarray*}
\begin{eqnarray*}
{P(\o)}^{(12)} & = & 273646526 - 761825992\gamma + 377695696\gamma^{2} + 89195520\gamma^{3}  \\
      &   &   + 45390432\gamma^{4} + 35391872\gamma^{5} + 28371968\gamma^{6}  \\
      &   &   + 15465472\gamma^{7} - 5471232\gamma^{8} - 28061696\gamma^{9} - 34398208\gamma^{10},
\end{eqnarray*}
\begin{eqnarray*}
{P(\o)}^{(13)} & = & -2326980998 + 7184044444\gamma - 4654680536\gamma^{2}  \\
      &   & - 637249840\gamma^{3} - 288190208\gamma^{4} - 259172928\gamma^{5}  \\
      &   & - 257515008\gamma^{6} - 227319040\gamma^{7} - 137749504\gamma^{8}  \\
      &   & + 15841280\gamma^{9} + 189190144\gamma^{10} + 240787456\gamma^{11},
\end{eqnarray*}
\begin{eqnarray*}
{P(\o)}^{(14)} & = & 20732504062 - 70283711216\gamma + 56240459224\gamma^{2}  \\
      &   &  + 2043271904\gamma^{3} + 731559936\gamma^{4} + 1283891648\gamma^{5}  \\
      &   &  + 1667097280\gamma^{6} + 1836321792\gamma^{7} + 1684263680\gamma^{8}  \\
      &   &  + 1097334784\gamma^{9} + 33075200\gamma^{10} - 1296547840\gamma^{11}  \\
      &   &  - 1704034304\gamma^{12},
\end{eqnarray*}
\begin{eqnarray*}
{P(\o)}^{(15)} & = & -192982729350 + 712495690468\gamma - 678540577872\gamma^{2}  \\
      &   &   + 43943729216\gamma^{3} + 15689764160\gamma^{4} - 248456832\gamma^{5}  \\
      &   &   - 6690721088\gamma^{6} - 9906840832\gamma^{7} - 11246331136\gamma^{8}  \\
      &   &   - 10574563328\gamma^{9} - 7651909632\gamma^{10} - 1206583296\gamma^{11}  \\
      &   &   + 9007038464\gamma^{12} + 12171673600\gamma^{13},
\end{eqnarray*}
\begin{eqnarray*}
{P(\o)}^{(16)} & = & 1871953992254 - 7474944990488\gamma + 8253899207808\gamma^{2}  \\
      &   &  - 1370040257024\gamma^{3} - 372411184768\gamma^{4}  \\
      &   &  - 94316090496\gamma^{5} - 14702702272\gamma^{6} + 15363239488\gamma^{7}  \\
      &   &  + 29522467968\gamma^{8} + 35529220864\gamma^{9} + 38287712768\gamma^{10}  \\
      &   &  + 40591720448\gamma^{11} + 15336308736\gamma^{12} - 63292702720\gamma^{13}  \\
      &   &  - 87636049920\gamma^{14},
\end{eqnarray*}
\begin{eqnarray*}
{P(\o)}^{(17)} & = & -18880288847750 + 81057814178860\gamma - 101782506154664\gamma^{2}  \\
      &   & + 27189637704816\gamma^{3} + 5335958139680\gamma^{4} + 1434188162432\gamma^{5}  \\
      &   & + 658007059712\gamma^{6} + 490637978560\gamma^{7} + 478988103296\gamma^{8}  \\
      &   & + 504910140800\gamma^{9} + 504004959232\gamma^{10} + 347157901824\gamma^{11}  \\
      &   & - 27525865472\gamma^{12} - 157623173120\gamma^{13} + 449134755840\gamma^{14}  \\
      &   & + 635361361920\gamma^{15} ,
\end{eqnarray*}
\begin{eqnarray*}
{P(\o)}^{(18)} & = & 197601208474238 - 907450604595520\gamma  \\
      &   & + 1276490317628872\gamma^{2} - 469854934268320\gamma^{3}  \\
      &   & - 57231645622432\gamma^{4} - 12599010163904\gamma^{5} - 8136201575040\gamma^{6}  \\
      &   & - 8857562356800\gamma^{7} - 10650704276864\gamma^{8} - 12583232475008\gamma^{9}  \\
      &   & - 14215451977344\gamma^{10} - 14614504454144\gamma^{11}  \\
      &   & - 11416776989184\gamma^{12} - 3717889196032\gamma^{13} + 1478858342400\gamma^{14}  \\
      &   & - 3214181007360\gamma^{15} - 4634400522240\gamma^{16} ,
\end{eqnarray*}
\begin{eqnarray*}
{P(\o)}^{(19)} & = & -2142184050841734 + 10475986286134644\gamma  \\
       &   &   - 16312769443915296\gamma^{2} + 7646757249760992\gamma^{3}  \\
       &   &   + 364024229452512\gamma^{4} + 20915048148288\gamma^{5}  \\
       &   &   + 60965998050816\gamma^{6} + 104195753619392\gamma^{7} + 143529329474688\gamma^{8}  \\
       &   &   + 180347215997184\gamma^{9} + 214040019465728\gamma^{10}  \\
       &   &   + 242779230361408\gamma^{11} + 251416724734720\gamma^{12}  \\
       &   &   + 201407890808064\gamma^{13} + 80080105439232\gamma^{14}  \\
       &   &   - 13205549875200\gamma^{15} + 23172002611200\gamma^{16}  \\
       &   &   + 33985603829760\gamma^{17},
\end{eqnarray*}
\begin{eqnarray*}
{P(\o)}^{(20)} & = & 24016181943732414 - 124577072070506344\gamma  \\
      &   &    + 212664085816703088\gamma^{2} - 121030264397636800\gamma^{3}  \\
      &   &    + 3207001016515744\gamma^{4} + 1926788775361664\gamma^{5}  \\
      &   &    - 97846534908224\gamma^{6} - 1011525246587968\gamma^{7}  \\
      &   &    - 1613872175868224\gamma^{8} - 2122232829566464\gamma^{9}  \\
      &   &    - 2582498914206400\gamma^{10} - 3029609305192704\gamma^{11}  \\
      &   &    - 3468160142157312\gamma^{12} - 3654216100258816\gamma^{13}  \\
      &   &    - 2984599157455872\gamma^{14} - 1274373807013888\gamma^{15}  \\
      &   &    + 114396518154240\gamma^{16} - 168139303157760\gamma^{17}  \\
      &   &    - 250420238745600\gamma^{18},
\end{eqnarray*}
\begin{eqnarray*}
{P(\o)}^{(21)} & = & -278028611833689478 + 1524422965551679164\gamma  \\
      &   &   - 2830011720336190520\gamma^{2} + 1893620809129623184\gamma^{3}  \\
      &   &   - 190312900454425856\gamma^{4} - 52429573013981632\gamma^{5}  \\
      &   &   - 5696486069346432\gamma^{6} + 8870271163894976\gamma^{7}  \\
      &   &   + 16508258449035520\gamma^{8} + 22318365235658944\gamma^{9}  \\
      &   &   + 27330506775020352\gamma^{10} + 32127595322607680\gamma^{11}  \\
      &   &   + 37750764697213632\gamma^{12} + 44585314916972224\gamma^{13}  \\
      &   &   + 48710458796935168\gamma^{14} + 40826850701992704\gamma^{15}  \\
      &   &   + 18137503276204032\gamma^{16} - 971272783134720\gamma^{17}  \\
      &   &   + 1227059169853440\gamma^{18} + 1853109766717440\gamma^{19},
\end{eqnarray*}
\begin{eqnarray*}
{P(\o)}^{(22)} & = & 3319156078802044158 - 19176747359879923600\gamma  \\
      &   &    + 38453690823403414456\gamma^{2} - 29561658453026305696\gamma^{3}  \\
      &   &    + 5155032487045587520\gamma^{4} + 942243974334873216\gamma^{5}  \\
      &   &    + 105439299259768768\gamma^{6} - 81111203950262400\gamma^{7}  \\
      &   &    - 160997366287047168\gamma^{8} - 215954557589465664\gamma^{9}  \\
      &   &    - 259399767952654784\gamma^{10} - 296956113372155968\gamma^{11}  \\
      &   &    - 341510419736224256\gamma^{12} - 415727829470043072\gamma^{13}  \\
      &   &    - 528023625074855232\gamma^{14} - 615174223548672512\gamma^{15}  \\
      &   &    - 534970609920598272\gamma^{16} - 244776032213663744\gamma^{17}  \\
      &   &    + 8131502895267840\gamma^{18} - 9000818866913280\gamma^{19}  \\
      &   &    - 13765958267043840\gamma^{20},
\end{eqnarray*}
\begin{eqnarray*}
{P(\o)}^{(23)} & = & -40811417293301014150 + 247769892998148929540\gamma  \\
      &   &    - 533551218840123871408\gamma^{2} + 463133218427313661568\gamma^{3}  \\
      &   &    - 114908031635611603008\gamma^{4} - 13026651600988443520\gamma^{5}  \\
      &   &    - 694835794636118976\gamma^{6} + 985945653624581504\gamma^{7}  \\
      &   &    + 1559657950629139840\gamma^{8} + 1927443716610738944\gamma^{9}  \\
      &   &    + 2162116688501650432\gamma^{10} + 2286453247614987008\gamma^{11}  \\
      &   &    + 2406004053525926464\gamma^{12} + 2793344171831971456\gamma^{13}  \\
      &   &    + 3867839365244115904\gamma^{14} + 5768757000308409472\gamma^{15}  \\
      &   &    + 7479458725939469568\gamma^{16} + 6842447109112948224\gamma^{17}  \\
      &   &    + 3214950387582763008\gamma^{18} - 67385809698816000\gamma^{19}  \\
      &   &    + 66326889832120320\gamma^{20} + 102618961627054080\gamma^{21},
\end{eqnarray*}
\begin{eqnarray*}
{P(\o)}^{(24)} & = & 516247012345341914942 - 3285119025877372180920\gamma  \\
      &   &    + 7559108520607773843488\gamma^{2} - 7308871611544450832064\gamma^{3}  \\
      &   &    + 2350981727669407843776\gamma^{4} + 118905993297229011008\gamma^{5}  \\
      &   &    - 16790780767539633792\gamma^{6} - 16613440672679879680\gamma^{7}  \\
      &   &    - 15548046461070182784\gamma^{8} - 15384986519988719936\gamma^{9}  \\
      &   &    - 14161472648995235136\gamma^{10} - 11276361716222164096\gamma^{11}  \\
      &   &    - 7324030172672791168\gamma^{12} - 4587460726218948608\gamma^{13}  \\
      &   &    - 8052405463773808896\gamma^{14} - 24820684069180820672\gamma^{15}  \\
      &   &    - 56960923428566366080\gamma^{16} - 88298755351002966272\gamma^{17}  \\
      &   &    - 86378467061276064256\gamma^{18} - 41668128760418795520\gamma^{19}  \\
      &   &    + 554175039327436800\gamma^{20} - 490786338216345600\gamma^{21}  \\
      &   &    - 767411365211013120\gamma^{22},
\end{eqnarray*}
\begin{eqnarray*}
{P(\o)}^{(25)} & = & -6711185258405244576646 + 44661059313146988290380\gamma  \\
      &   &    - 109331606843615382157384\gamma^{2} + 116480278944698068262320\gamma^{3}  \\
      &   &    - 46011745270359552516256\gamma^{4} + 440850798604678021632\gamma^{5}  \\
      &   &    + 792105834451739504064\gamma^{6} + 305738012971578476416\gamma^{7}  \\
      &   &    + 159590366941358300288\gamma^{8} + 95740210483655793728\gamma^{9}  \\
      &   &    + 30567600232164220480\gamma^{10} - 56955396275740956800\gamma^{11}  \\
      &   &    - 166253571659143480512\gamma^{12} - 281518130521880300608\gamma^{13}  \\
      &   &    - 362613423701499252928\gamma^{14} - 326561982884636510016\gamma^{15}  \\
      &   &    - 57663159522038980416\gamma^{16} + 474502223919197873792\gamma^{17}  \\
      &   &    + 1016510392155756708096\gamma^{18} + 1083954865492145854976\gamma^{19}  \\
      &   &    + 537244083313897897984\gamma^{20} - 4530850240533626880\gamma^{21}  \\
      &   &    + 3645203984752312320\gamma^{22} + 5755585239082598400\gamma^{23},
\end{eqnarray*}
\begin{eqnarray*}
{P(\o)}^{(26)} & = & 89574471680939133937534 - 622083509256483761778144\gamma  \\
      &   &    + 1613972832019449938327400\gamma^{2} - 1877838552918714034411104\gamma^{3}  \\
      &   &    + 879136906924104701402400\gamma^{4} - 60203145083669334396480\gamma^{5}  \\
      &   &    - 20498110028978700736576\gamma^{6} - 5129659282973757956032\gamma^{7}  \\
      &   &    - 1545427282000245352192\gamma^{8} - 81470942122801407680\gamma^{9}  \\
      &   &    + 1260555105289740810752\gamma^{10} + 2895967959719246612096\gamma^{11}  \\
      &   &    + 4862870780663281576640\gamma^{12} + 7053872326919183964416\gamma^{13}  \\
      &   &    + 9217493906833607526464\gamma^{14} + 10750408710797132958720\gamma^{15}  \\
      &   &    + 10329246189184106668160\gamma^{16} + 6079839402236030452352\gamma^{17}  \\
      &   &    - 2473100644141600758272\gamma^{18} - 11423977927106528212480\gamma^{19}  \\
      &   &    - 13587010293017989388288\gamma^{20} - 6925115206590254809088\gamma^{21}  \\
      &   &    + 36874116098389180416\gamma^{22} - 27166362328469864448\gamma^{23}  \\
      &   &    - 43282000997901139968\gamma^{24},
\end{eqnarray*}
\begin{eqnarray*}
{P(\o)}^{(27)} & = & -1226366187219563392423046 + 8871463075992006738446996\gamma  \\
      &   &    - 24310664899996326040840960\gamma^{2} + 30660793724129104401682976\gamma^{3}  \\
      &   &    - 16590293346041070916979232\gamma^{4}+ 2101735937627830336699008\gamma^{5}  \\
      &   &    + 414485759213945048620480\gamma^{6} + 69997075687476971109760\gamma^{7}  \\
      &   & + 9796719680443211190976\gamma^{8} - 12989214479805658674880\gamma^{9} \\ 
      &   & - 34255224236967502762880\gamma^{10} - 59677881433243896747264\gamma^{11} \\
      &   & - 89567316898510113800576\gamma^{12} - 122844647469002814372800\gamma^{13}  \\
      &   &    - 158096743387017593763520\gamma^{14}- 192783430108635063948928\gamma^{15}  \\
      &   &    - 218698994059862664719168\gamma^{16} - 214808919929359674024448\gamma^{17}  \\
      &   &    - 150396704456781735005888\gamma^{18}- 17357306803247632596736\gamma^{19} \\ 
      &   & + 125070331075601599340800\gamma^{20} + 170681832575907668260864\gamma^{21} \\
      &   & + 89515662319906963062784\gamma^{22} - 299007080543601819648\gamma^{23}  \\
      &   &  + 203092466220920733696\gamma^{24}+ 326279699830331670528\gamma^{25}.
\end{eqnarray*}

\begin{thebibliography}{1}

\bibitem{Evans-93}
J. W. Evans, {Rev. Mod. Phys.} {\bf 65}, 1281 (1993).

\bibitem{Privman-Nielaba-92}
V. Privman and P. Nielaba, {Europhys. Lett.} {\bf 18}, 673 (1992).

\bibitem{Nielaba-Privman-92}
P. Nielaba and V. Privman, {Mod. Phys. Lett. B} {\bf 6}, 533 (1992).

\bibitem{Wang-group-93}
J.~-S.~Wang, P. Nielaba, and V. Privman,
{Mod. Phys. Lett. B} {\bf 7}, 189 (1993).

\bibitem{Baram-Kutasov-89}
A. Baram and D. Kutasov, {J. Phys. A: Math. Gen.} {\bf 22}, L251 (1989).

\bibitem{Dickman-Wang-Jensen-91}
R. Dickman, J. -S. Wang, and I. Jensen, 
{J. Chem. Phys.} {\bf 94}, 8252 (1991).

\bibitem{Oliveira-group-92}
M. J.~de Oliveira and T. Tom\'e, {Phys. Rev. A} {\bf 46}, 6294
(1992).

\bibitem{Bonnier-group-93}
B. Bonnier, M. Hontebeyrie and C. Meyers, {Physica A} {\bf 198},
1 (1993).

\bibitem{Baram-Fixman-95}
A. Baram and M. Fixman, {J. Chem. Phys.} {\bf 103}, 1929 (1995).

\bibitem{Gan-Wang-96}
C. K. Gan and J. -S. Wang, {J. Phys. A: Math. Gen} {\bf 29}, L177 (1996).

\bibitem{Martin-74}
J. L. Martin, in {\it Phase Transitions and Critical Phenomena} {\bf 3},
ed. by C. Domb and M. S. Green (New York: Academic) p. 97.

\bibitem{Grynberg-Stinchcombe-95}
M.~D. Grynberg and R. B. Stinchcombe, {Phys. Rev. Lett.} {\bf 74},
1242 (1995).

\bibitem{Avraham-group-90}
D.~b.~-Avraham, M. A. Burschka, and C. R. Doering,
{J. Stat. Phys.} {\bf 60}, 695 (1990).

\bibitem{Lushnikov-87}
A.~A.~Lushnikov,
{Phys. Lett. A} {\bf 120}, 135 (1987).

\bibitem{Spouge-88}
J.~L.~Spouge,
{Phys. Rev. Lett.} {\bf 60}, 871 (1988).

\bibitem{Balding-group-88}
D.~Balding, P.~Clifford, and N.~J.~B.~Green,
{Phys. Lett. A} {\bf 126}, 481 (1988).

\bibitem{Baker-61}
G.~A.~Baker,~Jr., 
{Phys. Rev.} {\bf 124}, 768 (1961).

\bibitem{Hunter-Baker-73}
D.~L.~Hunter and G.~A.~Baker,~Jr., 
{Phys. Rev. B} {\bf 7}, 3346 (1973).

\bibitem{Baker-Hunter-73}
G.~A.~Baker,~Jr. and D.~L.~Hunter,
{Phys. Rev. B} {\bf 7}, 3377 (1973).

\bibitem{Brosilow-group-91}
B. J. Brosilow, R. M. Ziff, and R.~D.~Vigil,
{Phys. Rev. A}, {\bf 43}, 631 (1991).

\bibitem{Wang-94}
J.~-S.~Wang, {Inter. J. Mod. Phys. C} {\bf 5}, 707 (1994).


\bibitem{Privman-Barma-92}
V. Privman and M. Barma,
{J. Chem. Phys.} {\bf 97}, 6714 (1992).


\bibitem{Jensen-Dickman-93}
I. Jensen and R. Dickman,
{J. Stat. Phys.} {\bf 71}, 89 (1993).

\bibitem{Privman-private}
V. Privman (private communication).

\bibitem{Song-Poland-92}
S. Song and D. Poland,
{J. Phys. A} {\bf 25}, 3913 (1992).
\end{thebibliography}
\end{document}